\documentclass[3p,12pt]{elsarticle}

\usepackage{lineno,hyperref}
\usepackage{graphicx,subfloat,enumerate}
\usepackage{amssymb}
\usepackage{bm,amsmath}
\usepackage[caption=false]{subfig}
\usepackage{caption,color}
\usepackage{subeqnarray}
\usepackage{multirow}
\usepackage{lscape}
\usepackage{rotating}
\usepackage{graphics}
\usepackage{epstopdf}

\usepackage{graphicx} 
\usepackage{enumerate}
\usepackage{amsmath}
\usepackage{esint}
\usepackage{float}
\usepackage{booktabs}

\newcommand{\leir}[1]{{\leavevmode\color{red}#1}}

\biboptions{sort&compress}

\journal{XXX}

\begin{document}

\begin{frontmatter}

\title{Multiscale simulation of rarefied gas flows in Divertor Tokamak Test facility}


\author{Wei Li, Yanbing Zhang, Jianan Zeng}
\author{Lei Wu\corref{mycorrespondingauthor}}
\cortext[mycorrespondingauthor]{Corresponding author}
\ead{wul@sustech.edu.cn}

\address{Department of Mechanics and Aerospace Engineering, Southern University of Science and Technology, Shenzhen 518055, China}

\begin{abstract}
Simulating gas flow within the divertor, which is a crucial component in nuclear fusion reactors, is essential for assessing and enhancing its design and performance. Traditional methods, such as the direct simulation Monte Carlo and the discrete velocity method, often fall short in efficiency for these simulations. In this study, we utilize the general synthetic iterative scheme to simulate a simplified Tokamak divertor model, demonstrating its fast convergence and asymptotic-preserving properties in complex three-dimensional scenarios. A conservative estimate of speedup by three orders of magnitude is achieved by the general synthetic iterative scheme when compared to the direct simulation Monte Carlo method.
We further investigate the relationship between pumping efficiency and factors like temperature, absorptivity, and the Knudsen number, providing valuable insights to guide the design and optimization of divertor structures.
\end{abstract}

\begin{keyword}
vacuum pump, rarefied gas flow, general synthetic iterative scheme
\end{keyword}

\end{frontmatter}


\section{Introduction}


Nuclear fusion, a reaction process with immense energy potential, is considered a promising source of future clean energy. Divertors are essential components in fusion reactors, which enhance the efficiency and sustainability of fusion reactions by reducing energy loss and impurity accumulation.
For example, the Divertor Tokamak Test (DTT) facility in Europe aims to conduct scaled experiments to develop divertor solutions compatible with the anticipated physical conditions and technological environment of the DEMO reactor. The pumping rate is one of the critical factors in the design of divertor~\cite{tantos2023dtt,simulationJET,simulationJT60}, and numerical methods are used to determine optimal pumping port configurations~\cite{DTT2015,DTT2019}. 
However, the complex structure of divertor results in a wide range of Knudsen numbers (Kn, the ratio of mean molecular free path $\lambda$ to the characteristic flow length $L$), and poses significant challenges in the numerical simulations, as the gas flow should be described by the Boltzmann equation rather than the traditional Navier-Stokes equations. 

The Boltzmann equation can be solved by the stochastic direct simulation Monte Carlo (DSMC) method~\cite{bird1994molecular} and deterministic discrete velocity method (DVM)~\cite{Aristov2001}. The DSMC uses simulation particles to mimic the streaming and collision of real gas molecules, and there are only a few simulation particles in each spatial cell. Therefore, it has become the prevailing method to simulate the rarefied gas dynamics as the usage of computer memory is acceptable. However, because the streaming and collision are splitted, the cell size and time step must be smaller than the mean free path and mean collision time of gas molecules, respectively, rending the DSMC extremely time-consuming in simulating near-continuum flows. Worse still, since the statistical averaging is needed, the DSMC is slow in resolving small and/or transit flow fields. As a consequence, in the simulation of DTT particle exhaust~\cite{tantos2023dtt}, 40 million spatial cells and 0.688 million CPU core hours are required to find the steady state, making the optimization of divertor difficult.


In DVM, in addition to the spatial discretization, the molecular velocity space is also discretized. Since each physical cell contains thousands of discrete velocity points, the computer memory requirement can be hundreds times greater than that of the DSMC. However, due to its deterministic nature, the statistic averaging process is eliminated, making it faster than the DSMC in simulating low-speed and/or transit flows.  

Early versions of DVM also separate the streaming and collision processes, leading to large numerical dissipation similar to the DSMC.
In the past decades, significant progresses are made by Chinese scholars to eliminate these deficiencies and boost the simulation efficiency by several orders of magnitude. For example, the implicit unified gas-kinetic scheme (UGKS)~\cite{zhuyajun2016} and the generalized synthetic iteration scheme (GSIS)~\cite{SuArXiv2019,Su2021CMAME,Zeng2023CaF} have been proposed and applied to challenging multiscale engineering applications. In UGKS, the analytic solution of the kinetic equation is utilized to simultaneously handle streaming and collision, and the limitation of spatial cell size is removed. In GSIS, the traditional DVM is used to solve the kinetic equations, together with the macroscopic synthetic equations to empower the fast convergence and asymptotic-preserving characteristics. As a consequence, steady-state solutions can be obtained within dozens of iterations across the entire range of gas rarefaction. Compared to the UGKS and its variants \cite{Zhu2019JCP,guo2013discrete}, 
the flux evaluation in GSIS is much simpler, and more importantly, the macroscopic synthetic equations are solved to the steady state (while that in UGKS is only solved for one-time step), which significantly enhance the exchange of fluid information in the whole computational domain. As a result, if the same numbers of spatial cell and discrete velocity are used, the GSIS is faster~\cite{Zhang2024}. 

The aim of the present work is twofold. First, to use the GSIS to show that rarefied gas flows in divertor  can be efficiently simulated. Second, to perform numerical simulations over a wide range of Knudsen number, temperature, and surface absorptivity to test the performance of the divertor.

The rest of the paper is structured as follows: Section 2 introduces the divertor geometry and flow configuration. Section 3 details the kinetic model and numerical method used. Section 4 validates the convergence of both spatial and velocity space discretizations. Section 5 systematically analyzes the impact of Knudsen number, absorptivity, and temperature on pumping speed. Finally, conclusions are presented in Section 6.

 \section{Statement of the problem}

\begin{figure}[t]
    \centering
    \subfloat[Origin geometry]{\includegraphics[trim={0 0 0 0},clip, width=0.5\linewidth]{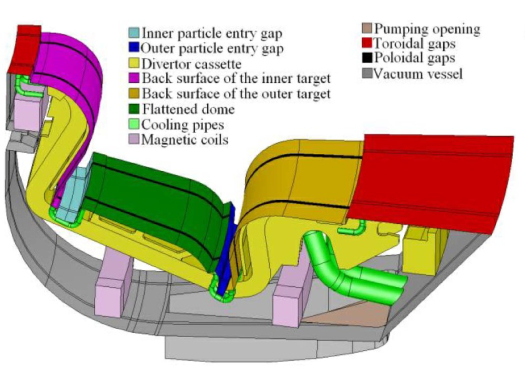}}\\
    \subfloat[Top view]{\includegraphics[trim={0 0 0 40},clip, width=0.55\linewidth]{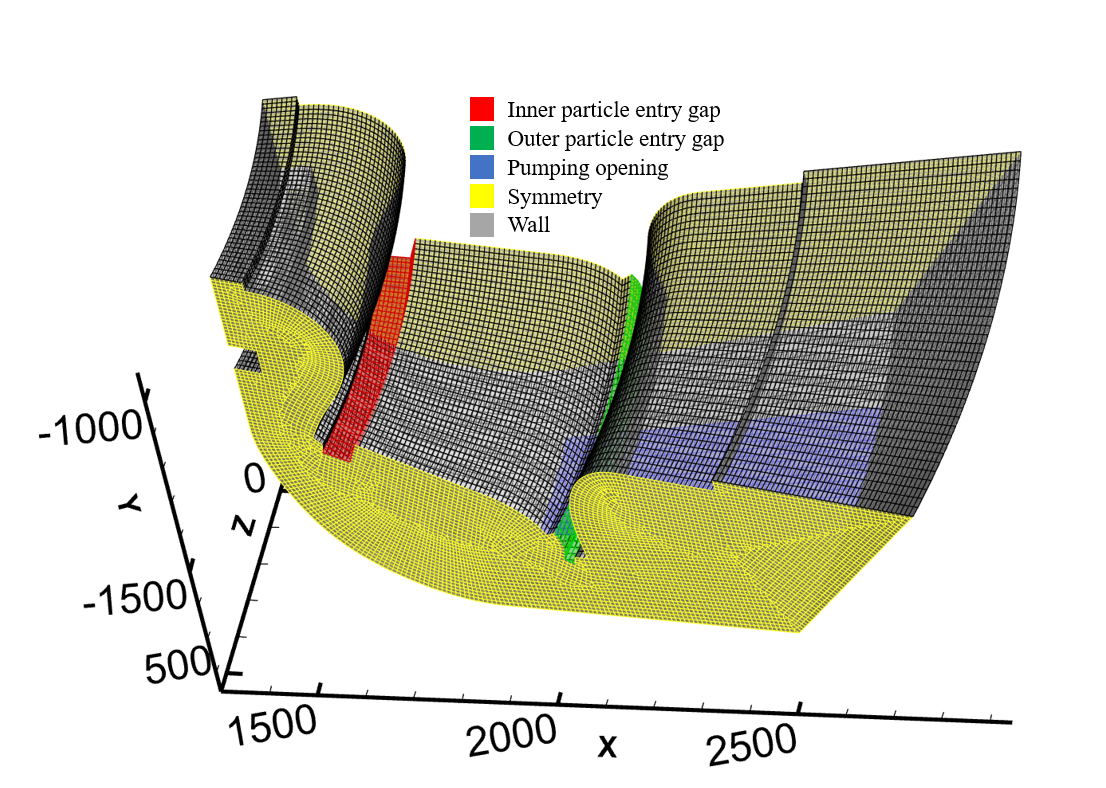}}
    \subfloat[Bottom view]{\includegraphics[trim={30 60 50 100},clip, width=0.44\linewidth]{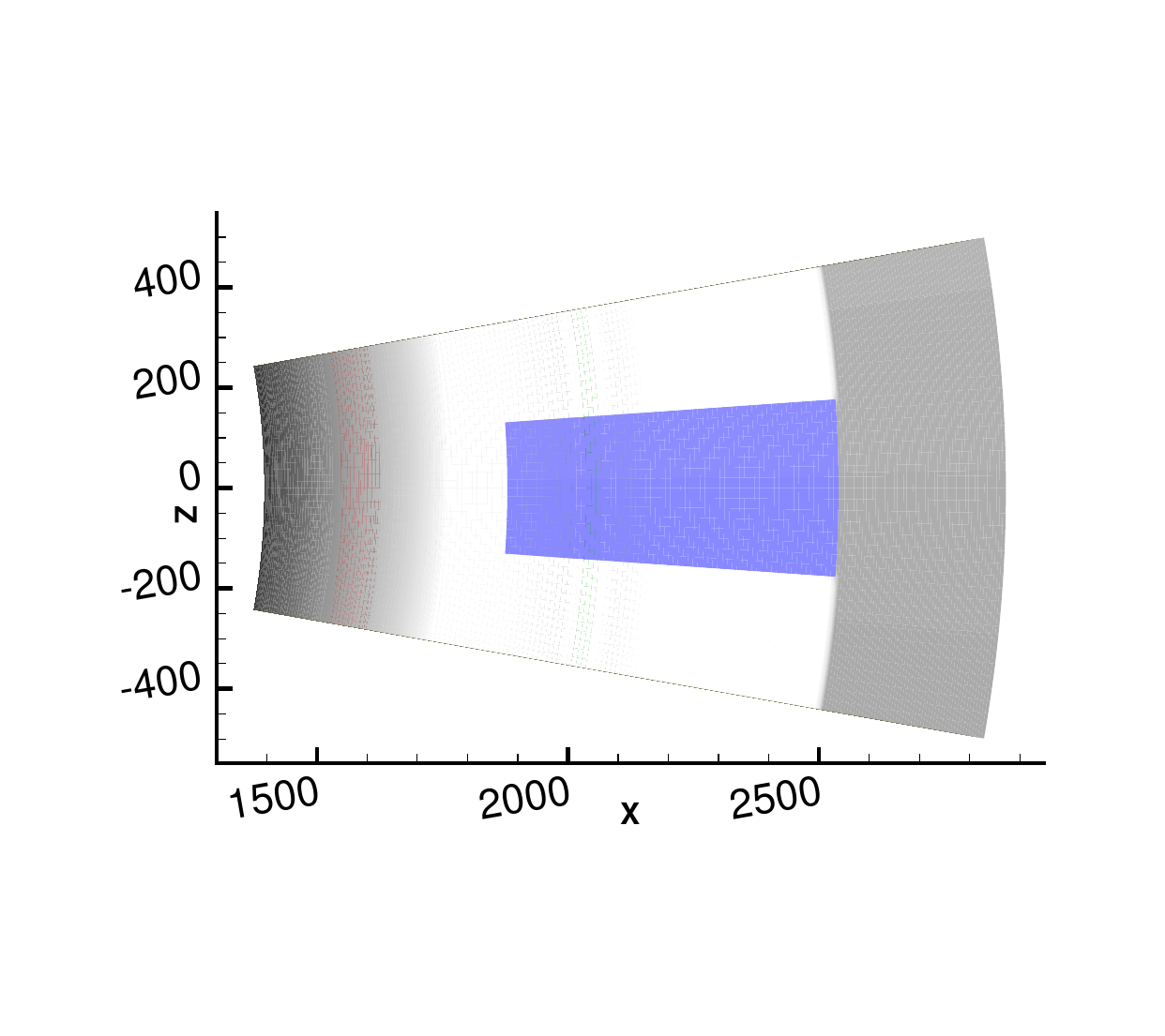}} 
    \caption{
    (a) Original geometry of the divertor obtained from Ref.~\cite{tantos2023dtt}. (b,c) The top and bottom views of the simplified geometry (Dimensions in mm).
    } 
    \label{mesh}
\end{figure}


The computational paradigm employs the latest DTT divertor \cite{DTT2019}, simulating 20° toroidal sector in Fig.~\ref{mesh}(a), out of the 18 sectors in total that form the entire DTT divertor ring. The origin model incorporates the inner surface of the vacuum vessel, the pumping duct, the poloidal gap, the toroidal interstice between the target and the divertor, as well as the toroidal magnetic coils and coolant conduits integrated within the divertor assembly. In this paper, we consider a simplified version of the divertor in Fig.~\ref{mesh}(b, c), without considering the magnetic coils, the cooling pipes and the toroidal/poloidal gaps.

Neutral deuterium gas enters the simulation domain through the inner and outer entry gaps in Fig.~\ref{mesh}(b).
These inlets are modeled as surfaces with specific capture coefficients~\cite{tantos2023dtt}, that is,
approximately 35\% and 65\% of the deuterium gas enters through the inner and outer entry gap, respectively. The flux of deuterium entering the divertor can be expressed as:
\begin{equation}\label{flow}
    \Phi_{in} = \frac{A}{4}\frac{\rho}{m}\sqrt{\frac{8k_{B}T}{\pi m}},
\end{equation}
where A is the area of the inner and the outer entry gap, $k_B$ is the Boltzmann constant;  $ \rho$, $m$, and $T$ are the density, molecular mass, and temperature of the gas, respectively. In typical operations, it is estimated in Ref.~\cite{tantos2023dtt} that the number densities at the inner entry gap is $ 1\times 10^{21}~\text{m}^{-3}$ and outer entry gap is $1.857\times 10^{21}~\text{m}^{-3}$.

The gas undergoes both inter-molecular collisions and interactions with the internal walls. Upon impacting with the walls, gas particles are reflected diffusely, where the wall temperature is maintained at 293~K. 
When gas molecules encounter the two lateral boundaries in Fig.~\ref{mesh}(b), they undergo specular reflection back into the simulation domain to mimic the toroidal symmetry~\cite{varoutis2019assessment}.

Beneath the divertor lies an absorption pump, which
is located in the middle of the bottom of the divertor, see the blue area in Fig.~\ref{mesh}(b,c).
Gas molecules have a probability of being absorbed, with a specified pump probability $\zeta$. That is, gas particles impinging on this surface have a probability $ \zeta $ of being absorbed and a probability $ 1-\zeta $ of being diffusely reflected back into the divertor region. 
In the DTT, it is estimated that the pump probability ranges from $ 0.05 \leq \zeta \leq 0.3 $.

In the design of the divertor, the efficacy of the absorption pump can be effectively gauged through the evaluation of parameters such as mass flow rate and pumping speed.
From Eq.~\eqref{flow}, we know that the inlet flux is determined by number density and temperature, while the outlet flux is primarily influenced by the outlet absorptivity. For a given set of density, temperature, and absorptivity $\zeta$, the mass flow rate and pumping speed of the divertor are uniquely determined.
This paper delves into the interplay between the Knudsen number, temperature, absorption rate, and the consequential impact on the mass flow rate of the absorption pump. By unraveling these relationships, valuable insights are gleaned to enhance the deflector's design and operational efficiency.

\section{Numerical method}\label{kinetic_equations}

In this section, the gas kinetic model, the velocity discretization, the gas-surface boundary condition, and the GSIS, are introduced.

\subsection{The kinetic model}

Kinetic model equations simplified from the Wang-Chang \& Uhlenbeck equation \cite{WangCS} are usually adopted in numerical simulations to describe the dynamics of molecular gas in the whole range of gas rarefaction. The model equation applied in this work is initially developed by Rykov~\cite{Rykov1978} and recently extended to reflect the proper relaxations of energy and heat-flux exchanges between translational and internal modes~\cite{li2021uncertainty}. Two velocity distribution functions (VDFs), $f_0(t, \bm{x}, \bm{\xi})$ and $f_1(t, \bm{x}, \bm{\xi})$, are used to describe the translational and internal states of gas molecules, where $t$ is the time, $\bm{x}=(x_1,x_2,x_3)$ is the spatial coordinate, and $\bm{\xi}=(\xi_1,\xi_2,\xi_3)$ is the molecular velocity. We assume the internal degrees of freedom is $d_r$. 
The macroscopic quantities, such as the density $\rho$, flow velocity $\bm{u}$, deviatoric stress $\bm{\sigma}$, translational and rotational temperature $T_t$ and $T_r$, translational and rotational heat flux $\bm{q}_t$ and $\bm{q}_r$, are obtained by taking moments of VDFs $f_0$ and $f_1$:
\begin{equation}\label{eq:getmoment}
    \begin{aligned}
    \left(\rho,~\rho\bm{u},~\bm{\sigma},~\frac{3}{2}\rho RT_t,~\bm{q}_{t}\right)&=\int\left(1,~\bm{\xi},~\bm{c}\bm{c}-\frac{c^2}{3}\mathrm{I},~\frac{c^2}{2},~\frac{c^2}{2}\bm{c}
        \right) f_0 \mathrm{d}\bm{\xi},\\
        \left(\frac{d_r}{2}\rho RT_r,~\bm{q}_{r}\right)&=\int\left(1,~\bm{c}\right)f_1\mathrm{d}\bm{\xi},
    \end{aligned}
\end{equation}
where $\bm{c}=\bm{\xi}-\bm{u}$ is the peculiar velocity and $\textbf{I}$ is the $3\times 3$ identity matrix. 
The total temperature $T$ is defined as the equilibrium temperature between the translational and internal modes: $T=(3T_t+d_rT_r)/(3+d_r)$. The pressure related to the translational motion is $p_t=\rho R T_t$, while the total pressure is $p=\rho RT$, with $R=k_B/m$ being the gas constant.

In the absence of an external force, the evolution of VDFs is governed by the following kinetic equations:
\begin{equation}\label{general_model}
    \begin{aligned}
        &\frac{\partial f_0}{\partial t}+\bm{\xi}\cdot \nabla f_0 = \frac{g_{0t}-f_0}{\tau}+\frac{g_{0r}-g_{0t}}{Z_r\tau}, \\
        & \frac{\partial f_1}{\partial t}+\bm{\xi}\cdot \nabla f_1 = \frac{g_{1t}-f_1}{\tau}+\frac{g_{1r}-g_{1t}}{Z_r\tau}, 
    \end{aligned}
\end{equation}
where terms in the left (right)-hand-side describe the streaming (collision); $\tau$ and ${Z}_{r}\tau$ are the elastic and inelastic collision characteristic time, respectively, with ${Z}_{r}$ being the rotational collision number. The elastic collision conserves the kinetic energy, while the inelastic collision exchanges the translational and rotational energies. Here we choose $Z_r={10}/{3}$ and the collision time is determined by 
\begin{equation}
    \tau=\frac{\mu}{p_t},
\end{equation}
where $\mu$ is the shear viscosity. The power-law intermolecular potential is considered, so that the viscosity can be expressed as 
$
\mu(T_t)=\mu(T_0)\left({T_t}/{T_0}\right)^{\omega}$,
with $\omega$ the viscosity index and $T_0$ the reference temperature. For deuterium gas we choose $\omega=0.74$. 

The reference distribution functions are given by:
\begin{equation}
    \begin{aligned}
     g_{0t} &= \rho\left(\frac{1}{2\pi RT_t}\right)^{3/2}\exp\left(-\frac{c^2}{2RT_t}\right)\left[1+\frac{2\bm{q}_{t}\cdot\bm{c}}{15RT_tp_t}\left(\frac{c^2}{2RT_t}-\frac{5}{2}\right)\right],\\
        g_{0r}&= \rho\left(\frac{1}{2\pi RT}\right)^{3/2}\exp\left(-\frac{c^2}{2RT}\right)\left[1+\frac{2\bm{q}_{0}\cdot\bm{c}}{15RTp}\left(\frac{c^2}{2RT}-\frac{5}{2}\right)\right],\\
        g_{1t}&=\frac{d_r}{2}RT_rg_{0t} + \left(\frac{1}{2\pi RT_t}\right)^{3/2}\frac{\bm{q}_{r}\cdot\bm{c}}{RT_t}\exp\left(-\frac{c^2}{2RT_t}\right),\\
          g_{1r}&=\frac{d_r}{2}RTg_{0r} + \left(\frac{1}{2\pi RT}\right)^{3/2}\frac{\bm{q}_{1}\cdot\bm{c}}{RT}\exp\left(-\frac{c^2}{2RT}\right),
    \end{aligned}
\end{equation}
with $\bm{q}_{0},~\bm{q}_{1}$ being linear combinations of translational and internal heat fluxes:
\begin{equation}
    \begin{bmatrix} 
        \bm{q}_{0} \\ \bm{q}_{1} 
    \end{bmatrix}
    =
    \begin{bmatrix}		
        (2-3A_{tt})Z_r+1 & -3A_{tr}Z_r  \\		
        -A_{rt}Z_r & -A_{rr}Z_r+1 \\ 
    \end{bmatrix}
    \begin{bmatrix} 
    \bm{q}_{t} \\ \bm{q}_{r} 
    \end{bmatrix},
\end{equation}
where $\bm{A}=[A_{tt},A_{tr},A_{rt},A_{rr}]$ is determined by the relaxation rates of heat flux. In this paper, $ A_{tt} = 0.786, A_{tr} = -0.201, A_{rt} = -0.056 $, and $ A_{rr} = 0.842 $, as extracted from the DSMC~\cite{li2021uncertainty}. Thus, in the continuum flow regime, the transitional and rotational thermal conductivities $\kappa_t$ and $\kappa_r$ can be derived from the kinetic model by the Chapman-Enskog expansion as~\citep{li2021uncertainty}:
\begin{equation}\label{eq:kappaAnumber}
	\left[ 
      \begin{array}{ccc} 
        \kappa_t \\ \kappa_r 
      \end{array}
    \right]
	= \frac{\mu}{2}
	\left[ 
      \begin{array}{ccc} 
        A_{tt} & A_{tr} \\ A_{rt} & A_{rr} 
      \end{array}
    \right]^{-1}
    \left[ 
      \begin{array}{ccc} 
        5 \\ d_r 
      \end{array}
    \right].
\end{equation}

Note that the collision terms in Eq.~\eqref{general_model} becomes zero when the equilibrium state is reached, i.e., when the VDFs take the following form:
\begin{equation}\label{equilibirumVDF}
\begin{aligned}
    &f_0^{eq}(\rho,\bm{u},T)= \rho\left(\frac{1}{2\pi RT}\right)^{3/2}\exp\left(-\frac{c^2}{2RT}\right), \\
    &f_1^{eq}(\rho,\bm{u},T)= \frac{d_r}{2}RT f_0^{eq}(\rho,\bm{u},T).
\end{aligned}
\end{equation}

\subsection{The discretization of velocity space and quadrature}

In DVM, the continuous velocity space $\bm{\xi}$ is discretized, where the way of discretization and the corresponding quadrature significantly affect the numerical accuracy and efficiency. Here, three types of velocity discretization are considered. 
\begin{itemize}
    \item The Newton-Cotes quadrature with uniform discretization of the truncated velocity space. 
    Since after the normalization of the molecular velocity by $\sqrt{RT}$, the equilibrium VDF is close to $\exp(-\xi^2/2)$, the velocity space is truncated to the region of $[a,b]=[-5\sqrt{RT},5\sqrt{RT}]$, and the quadrature of the function $g$ (e.g., the product of VDF and velocity moments) can be approximated as
\begin{equation}
    \int_{a}^{b} g(\xi)d\xi \approx \sum\limits_{i=1}^N {g(\xi_i)\omega_i},
\end{equation}
where the integration point $\xi_i=a + h(i-1)$, with $ h = (b-a)/{(N-1)}$ and $i=1,2,\cdots, N$;  $\omega_i$ is the i-th weight, which can be set to $h$ when the trapezoidal rule is applied. 
The cross product method can be applied when evaluating three-dimensional integration. 

\item The Gauss-Hermite quadrature. 
\begin{equation}
     \int_{-\infty}^{+\infty} g(\xi)d\xi= \int_{-\infty}^{+\infty} \exp(-\xi^2)\frac{g(\xi)}{\exp(-\xi^2)}d\xi 
     \approx\sum\limits_{i=1}^N g(\xi_i)\frac{\omega_i}{\exp(-\xi_i^2)},
\end{equation}
where $\xi_i$ and $\omega_i$ are the nodes and weights of the Gauss-Hermite quadrature. Since this quadrature has the highest algebraic accuracy, i.e., the integration error is zero when ${g(\xi)}{\exp(\xi^2)}$ is a polynomial of $\xi$ with order less than or equal to $2N-1$, the Gauss-Hermite quadrature can effectively reduce the number of discretized velocity for low-speed near-equilibrium flows~\cite{ambrucs2012high,shi2021accuracy}.
The cross product method can be applied when evaluating three-dimensional integration. 



\item When the cylindrical coordinates are used, the velocity points in the xz plane are transformed from $\xi_x$ and $\xi_z$ to $\xi_r$ and ${\theta}$. In the $\xi_y$ direction, the distribution function is defined on $(0, +\infty)$, and we employ the Gauss-Laguerre  quadrature:
\begin{equation}
\begin{aligned}
  \int_0^\infty  g(\xi_r) {\xi_r} d{\xi_r}= & \int_0^\infty  \frac{g(\xi_r) }{\exp(-\xi_r^2)} {\xi_r} \exp(-\xi_r^2) d{\xi_r}\\
  \overset{x=\xi^2_r}{=}&\frac{1}{2}  \int_0^\infty  \frac{g(\sqrt{x}) }{\exp(-x)}  \exp(-x) d{x}\\
  \approx&\sum_{i=1}^{N_r} \frac{1}{2} \frac{g(\sqrt{x_i}) }{\exp(-x_i)}\omega_i
  =\sum_{i=1}^{N_r} \frac{1}{2} \frac{g(\xi_i) }{\exp(-\xi^2_i)}\omega_i,
\end{aligned}
\end{equation}
where $x_i$ and $\omega_i$ are the nodes and weights in the Gauss-Laguerre quadrature, and $\xi_i=\sqrt{x_i}$.
In the $\xi_y$ direction, the Gauss-Hermite quadrature of order $N_y$ is used, while in the $\theta$ direction, the Newton-Cotes quadrature is applied with $N_
\theta$ uniform sections. Eventually, the integral in the cylindrical coordinates are:
\begin{equation}  
\begin{aligned}
  \iiint_{\bm{\xi}} {g d\bm{\xi}} & 
  = \int_0^{2\pi} {\int_0^\infty  {\int_{ - \infty }^\infty 
  g(\xi_r,\theta, \xi_{y}) {\xi_r}d{\xi_y}d{\xi_r}d{{\theta} }} }   \\
  &\approx\frac{2\pi}{N_\theta} \sum\limits_{j=1}^{N_\theta} \sum\limits_{k=1}^{N_y} \frac{\omega_k}{\exp(-\xi^2_k)} {\int_0^\infty  g(\xi_r,\theta_j, \xi_{k}) {\xi_r}d{\xi_r}} \\
 &  \approx\frac{\pi}{N_\theta} \sum_{i=1}^{N_r}\sum\limits_{j=1}^{N_\theta} \sum\limits_{k=1}^{N_y} g(\xi_i,\theta_{j}, \xi_k) \frac{\omega_k}{\exp(-\xi^2_k)} 
   \frac{\omega_i }{\exp(-\xi^2_i)}.
\end{aligned} 
\end{equation}

\end{itemize}

\subsection{The wall boundary conditions}

The Boltzmann kinetic equation describes the gas-gas interaction. To fully determine the rarefied gas flows, the gas-surface  boundary condition should be specified. In the present paper, the diffuse, absorption and toroidal symmetry boundary conditions will be considered. 

First, the diffuse boundary condition is applied at the wall, where gas molecules are reflected diffusely from the moving wall in thermodynamic equilibrium. The velocity distribution of gas molecules at the moving wall is given by: 
\begin{equation}
f_{0,\text{wall}} = \left\{
    \begin{array}{cc}
        f_{0,in}, &  \bm{n}\cdot\bm{\xi}\geq 0, \\
        f_0^{\text{eq}}(\rho_\text{wall},\bm{0},T_\text{wall}), & \bm{n}\cdot\bm{\xi} < 0, 
    \end{array}
    \right.
\end{equation}
where $ f_{0,\text{in}} $ is the VDF of gas molecules incident on the wall, $T_\text{wall}$ is the wall temperature, and the density $\rho_\text{wall}$ is determined by the non-penetration condition:
\begin{equation}
    \int_{\bm{n}\cdot\bm{\xi} \geq 0} \bm{n}\cdot \bm{\xi}  f_{0,\text{in}} \mathrm{d}\bm{\xi} = -\int_{\bm{n}\cdot\bm{\xi}  < 0} \bm{n}\cdot \bm{\xi} f_{0}^{\text{eq}}(\rho_\text{wall},\bm{0},T_\text{wall}) \mathrm{d}\bm{\xi}.
\end{equation}
Similarly, the boundary condition for the internal degree of freedom is given by
\begin{equation}
    f_{1,\text{wall}} = \left\{
    \begin{array}{cc}
        f_{1,in}, &  \bm{n}\cdot\bm{\xi}\geq 0, \\
        \frac{d_r}{2}RT_\text{wall} f_0^{\text{eq}}(\rho_\text{wall},\bm{0},T_\text{wall}), & \bm{n}\cdot\bm{\xi} < 0.
    \end{array}
    \right.
\end{equation}

Second, the absorption boundary condition is applied at the absorption pump, i.e., the blue area in Fig.~\ref{mesh}(b,c).  
\begin{equation}
\begin{aligned}
&f_{0,\text{pump}} = \left\{
    \begin{array}{cc}
        f_{0,in}, &  \bm{n}\cdot\bm{\xi}\geq 0, \\
        (1-\zeta)f_0^{\text{eq}}(\rho_\text{wall},\bm{0},T_\text{wall}), & \bm{n}\cdot\bm{\xi} < 0, 
    \end{array}
    \right. \\
&f_{1,\text{pump}} = \left\{
    \begin{array}{cc}
        f_{1,in}, &  \bm{n}\cdot\bm{\xi}\geq 0, \\
        (1-\zeta)\frac{d_r}{2}RT_\text{wall} f_0^{\text{eq}}(\rho_\text{wall},\bm{0},T_\text{wall}), & \bm{n}\cdot\bm{\xi} < 0.
    \end{array}
    \right.
\end{aligned}
\end{equation}
The implementation of the diffuse and adsorption boundary conditions is elaborated in Ref.~\cite{Liu2024}.

\begin{figure}[h]
    \centering
\includegraphics[width=0.45\linewidth]{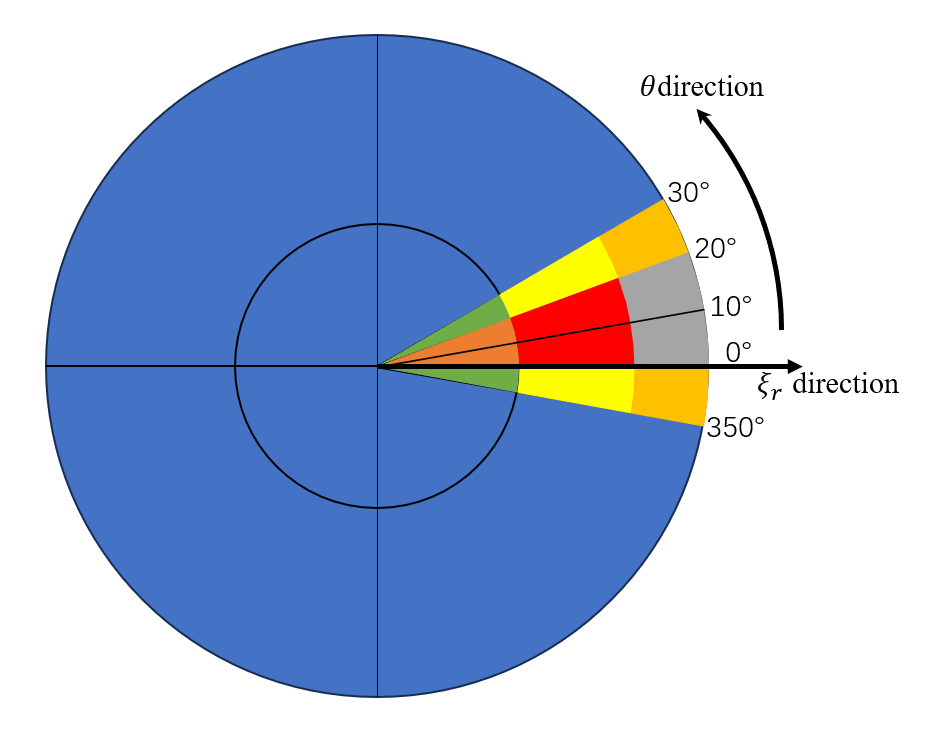}
    \caption{A schematic diagram of two-dimensional polar coordinate velocity discretization.}
    \label{symmetry}
\end{figure}

Third, the symmetry boundary condition is applied on the lateral walls in Fig.~\ref{mesh}(b), with the help of  cylindrical coordinates to avoid the interpolation in the velocity space. Since the $20^\circ$ sector is simulated, we discretize the velocity space in the ${\theta}$ direction uniformly, with $\Delta{\theta} = 10^{\circ}$, see Fig.~\ref{symmetry}. As shown in Fig.~\ref{mesh}(b,c), the front lateral wall projected in the $xz$ plane has the polar angle of -$10^\circ$, hence the symmetry boundary condition reads
\begin{equation}
 f(\xi_r,\theta,\xi_y)=f(\xi_r,-20^\circ-\theta,\xi_y), \quad
\theta\in[-10^\circ,170^\circ],
\end{equation}
while at the back lateral wall in Fig.~\ref{mesh}(b,c), with the polar angle being $10^{\circ}$ when projected to the $xz$ plane, the boundary condition reads
\begin{equation}
    f(\xi_r,\theta,\xi_y)=f(\xi_r,20^\circ-\theta,\xi_y), \quad
\theta\in[-170^\circ,10^\circ].
\end{equation}



\subsection{The GSIS}

After the velocity discretization, the kinetic equation can be solved by the iteration scheme in the finite volume framework~\cite{Zhang2024}. Since the two kinetic equations in Eq.~\eqref{general_model} have the same structure, to keep the presentation simple, we use the VDF $f$ to stand for either $f_0$ or $f_1$. Also, as here we focus on the spatial discretization, the index for the velocity discretization does not appear in the VDF.

Given the VDF and the macroscopic quantities in the n-th iteration step, the VDF at the $(n+1)$-th step can be obtained as
\begin{equation}\label{DVM0}
    \begin{aligned}[b]
        \frac{f_i^{n+1}-f_i^n}{\Delta t} + \frac{1}{V_i}\sum_{j\in N(i)} \xi_nf_{ij}^{n+1}S_{ij}=\frac{g^{n}_{i}-f^{n+1}_{i}}{\tau^{n}_i},
    \end{aligned}
\end{equation}
where 
\begin{equation}
        g = \left(1-\frac{1}{Z_r}\right)g_t+\frac{1}{Z_r}g_r.
\end{equation}
Here, $V_i$ is the volume of the $i$-th cell, $S_{ij}$ is the area of this cell and the $\Delta t$ is the time step.

This is the traditional DVM. Since the reference distribution function is evaluated at the $n$-th iteration step, it has large numerical dissipation at low spatial resolution. Therefore, in GSIS, the macroscopic synthetic equations are used to guide the evolution of the VDF. Firstly, by modifying Eq.~\eqref{DVM0}, we obtain the VDF at the intermediate iteration step $n+1/2$:
\begin{equation}
    \begin{aligned}[b]
        \frac{f_i^{n+1/2}-f_i^n}{\Delta t} + \frac{1}{V_i}\sum_{j\in N(i)} \xi_nf_{ij}^{n+1/2}S_{ij}=\frac{g^{n}_{i}-f^{n+1/2}_{i}}{\tau^{n}_i}.
    \end{aligned}
\end{equation}
Secondly, we solve the following macroscopic synthetic equations to get the macroscopic quantities at the $(n+1)$-th iteration step:
\begin{equation}\label{eq:macroscopic_equation_2}
	\begin{aligned}
		\frac{\partial{\rho}}{\partial{t}} + \nabla\cdot\left(\rho\bm{u}\right)  &= 0, \\
		\frac{\partial}{\partial{t}}\left(\rho\bm{u}\right) + \nabla\cdot\left(\rho\bm{u}\bm{u}\right) + \nabla\cdot\bm{P} &= 0, \\
		\frac{\partial}{\partial{t}}\left(\rho e\right) + \nabla\cdot\left(\rho e\bm{u}\right) + \nabla\cdot\left(\bm{P}\cdot\bm{u}+\bm{q}_t+\bm{q}_r\right) &= 0, \\
        \frac{\partial}{\partial{t}}\left(\rho e_r\right) + \nabla\cdot\left(\rho e_r\bm{u}+\bm{q}_r\right) &= \frac{d_r\rho R}{2}\frac{T-T_r}{Z_r\tau}.
	\end{aligned}
\end{equation} 
Here, $e_r=d_rRT_r/2$ and $e=(3RT_t+u^2)/2+e_r$ are the specific total and internal energies, respectively; the pressure tensor $\bm{P}$ is given by $\bm{P} = p_t\mathrm{I} + \bm{\sigma}$.
For general rarefied flows, the constitutive relation of the macroscopic equation should not only include Newton's law and Fourier's law of viscosity and heat flux, but also consider the higher-order rarefied effect:
\begin{equation}\label{eq:full_constitutive}
    \begin{aligned}
        \bm{\sigma}^{n+1} &= \bm{\sigma}^{n+1}_{\text{NSF}}  + 
        \underbrace{ \int \left(\bm{c}\bm{c}-\frac{c^2}{3}\mathrm{I}\right)f^{n+1/2}_0 \mathrm{d}\bm{v} -\bm{\sigma}^{n+1/2}_{\text{NSF}} }_{\text{HoT}_{\bm{\sigma}}},
        \\
        \bm{q}^{n+1}_{t} &=  \bm{q}^{n+1}_{t,\text{NSF}} + \underbrace{ {\frac{1}{2}}\int \bm{c}c^2f^{n+1/2}_0 \mathrm{d}\bm{v} -\bm{q}^{n+1/2}_{t,\text{NSF}} }_{\text{HoT}_{\bm{q}_{t}}},
        \\
        \bm{q}^{n+1}_{r} &= \bm{q}^{n+1}_{r,\text{NSF}} + \underbrace { \int \bm{c}f^{n+1/2}_1 \mathrm{d}\bm{v}  -\bm{q}^{n+1/2}_{r,\text{NSF}} }_{\text{HoT}_{\bm{q}_{r}}},
    \end{aligned}
\end{equation}
where the Navier-Stokes constitutive relations are given as:
\begin{equation}\label{eq:NSF_constitutive}
    \begin{aligned}
        \bm{\sigma}_{\text{NSF}} &= -\mu \left(\nabla\bm{u}+\nabla\bm{u}^{\mathrm{T}}-\frac{2}{3}\nabla\cdot\bm{u}\mathrm{I}\right),\\
        \bm{q}_{t,\text{NSF}} &= -\kappa_t\nabla T_t,\\
        \bm{q}_{r,\text{NSF}} &= -\kappa_r\nabla T_r.
    \end{aligned}
\end{equation}
Thirdly, the VDF at the (n+1)-th step is given by:
\begin{equation}
\begin{aligned}
    &f_0^{n+1}=f_0^{n+1/2} - f_0^{eq}(\bm{W}^{n+\frac{1}{2}})  +f_0^{eq}(\bm{W}^{n+1}),\\
   &f_1^{n+1}=f_1^{n+1/2} - \frac{d_r}{2}RT^{n+1/2}_rf_0^{eq}(\bm{W}^{n+\frac{1}{2}})  +\frac{d_r}{2}RT^{n+1}_rf_0^{eq}(\bm{W}^{n+1}), 
\end{aligned}
\end{equation}
where $\bm{W}=\{\rho,\bm{u}, T_t\}$. The detailed implementation of the GSIS method can be found in our recent paper~\cite{Zhang2024}.

\section{Convergence test and numerical efficiency}\label{section_GSIS2}

Convergence tests in both spatial and velocity spaces are carried out in GSIS. The simulation parameters are adopted from Ref.~\cite{tantos2023dtt}: the number densities at the inner and outer entry gaps are $ 1\times 10^{21}~\text{m}^{-3}$ and $ 1.857\times 10^{21}~\text{m}^{-3}$, respectively. The gas temperature is maintained at 293~K. 
Using a reference length of 1 dm, the Knudsen number
\begin{equation}
     \text{Kn}=\frac{\lambda}{L}\equiv
     \frac{\mu(T_0)}{p_0L}\sqrt{\frac{\pi R T_0}{2}}
 \end{equation}
is 0.0426. Larger Kn can be obtained by proportionally reducing the two number densities at the inner and outer entry gaps. The absorptivity is $\zeta = 0.3$.

\begin{figure}[t]
    \centering
    \subfloat[]{\includegraphics[trim={50 50 70 200},clip, width=0.45\linewidth]{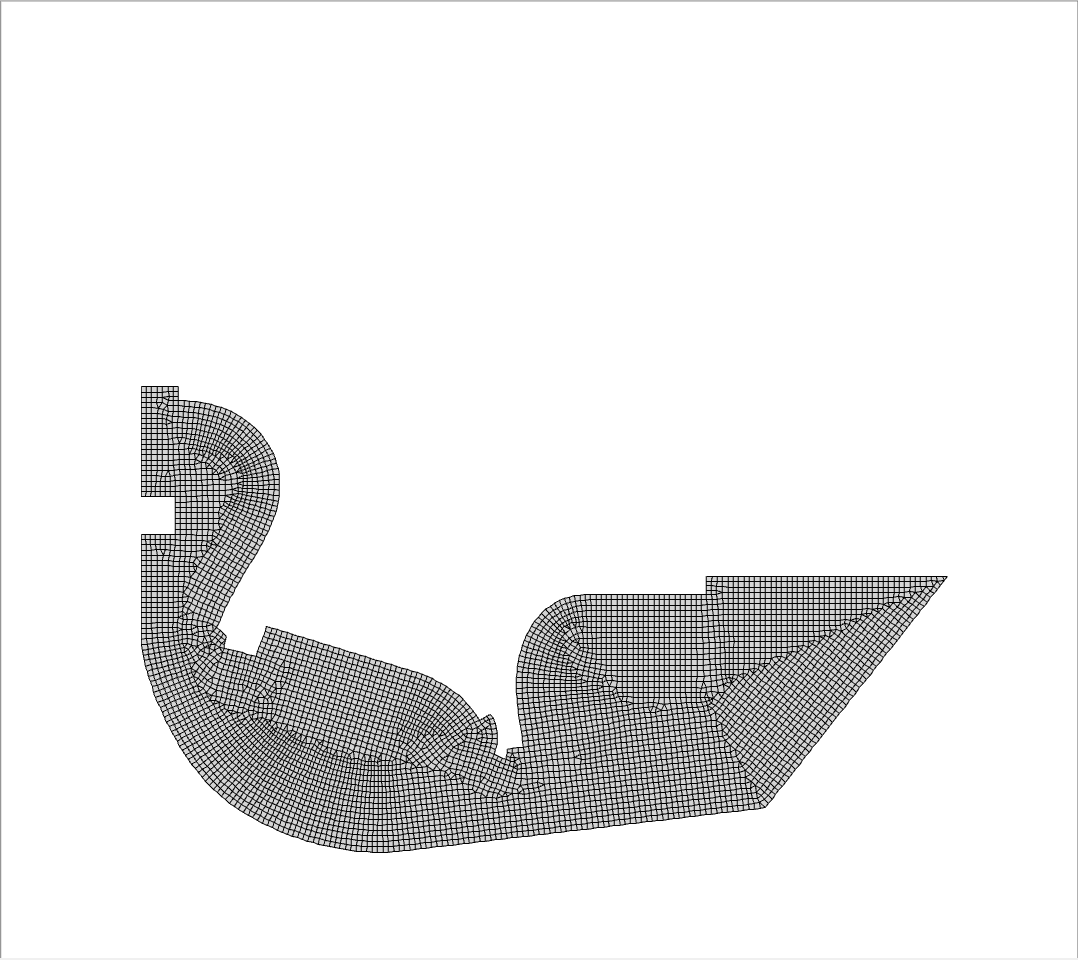}}
    \subfloat[]{\includegraphics[trim={50 50 70 200},clip,width=0.45\linewidth]{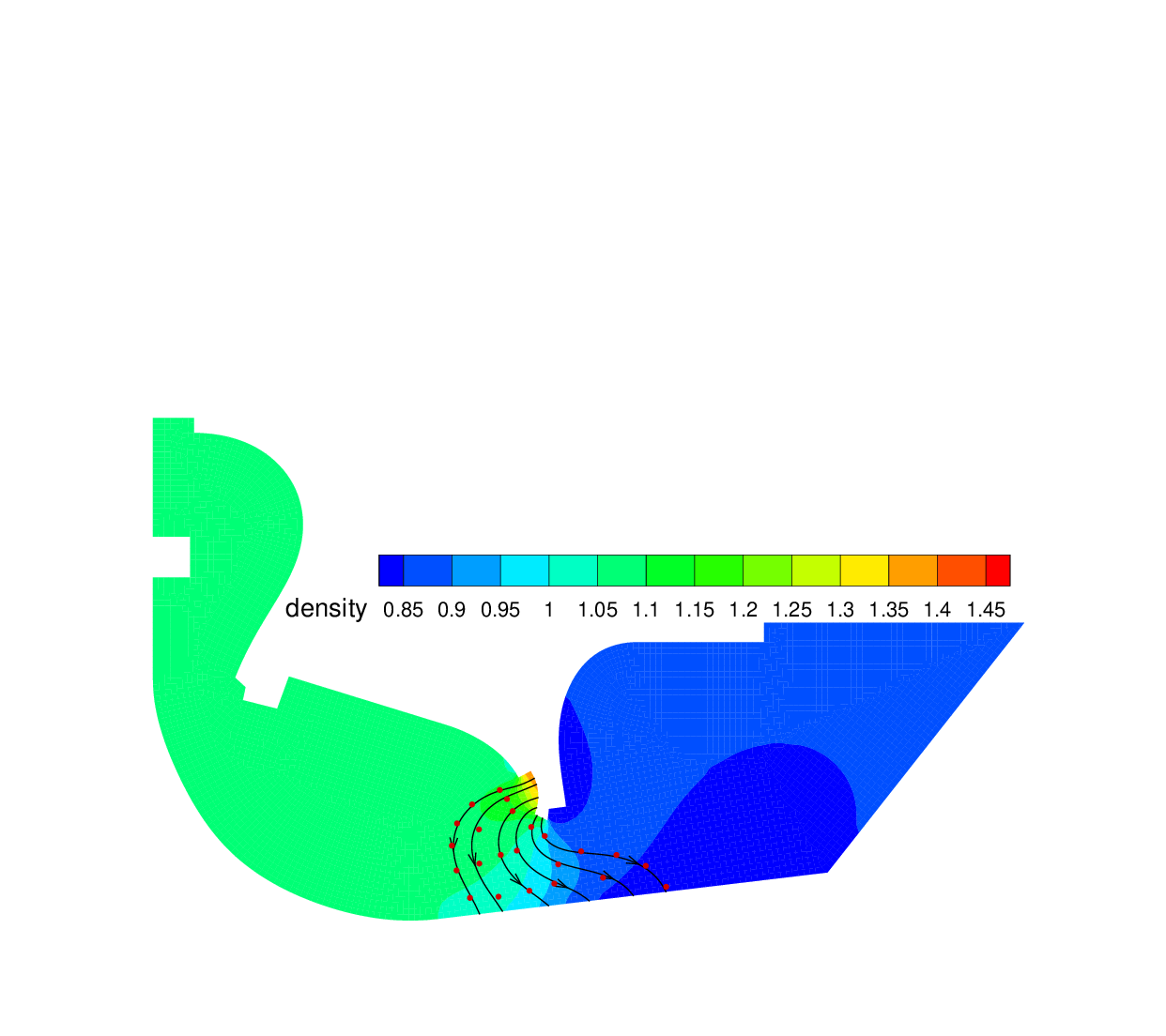}}\\
    \subfloat[]{\includegraphics[trim={50 50 70 220},clip,width=0.45\linewidth]{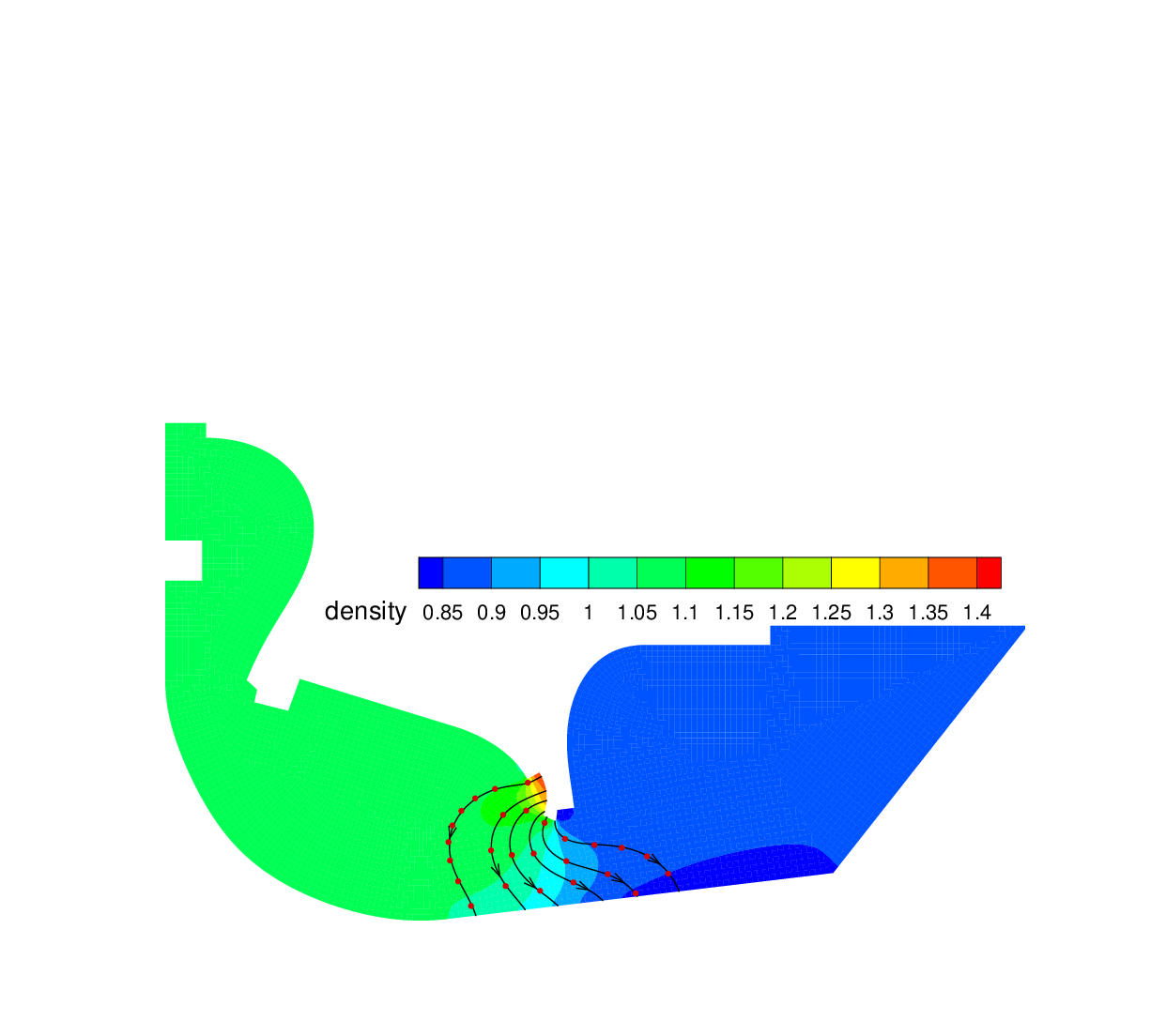}}
    \subfloat[]{\includegraphics[trim={50 50 70 220},clip, width=0.45\linewidth]{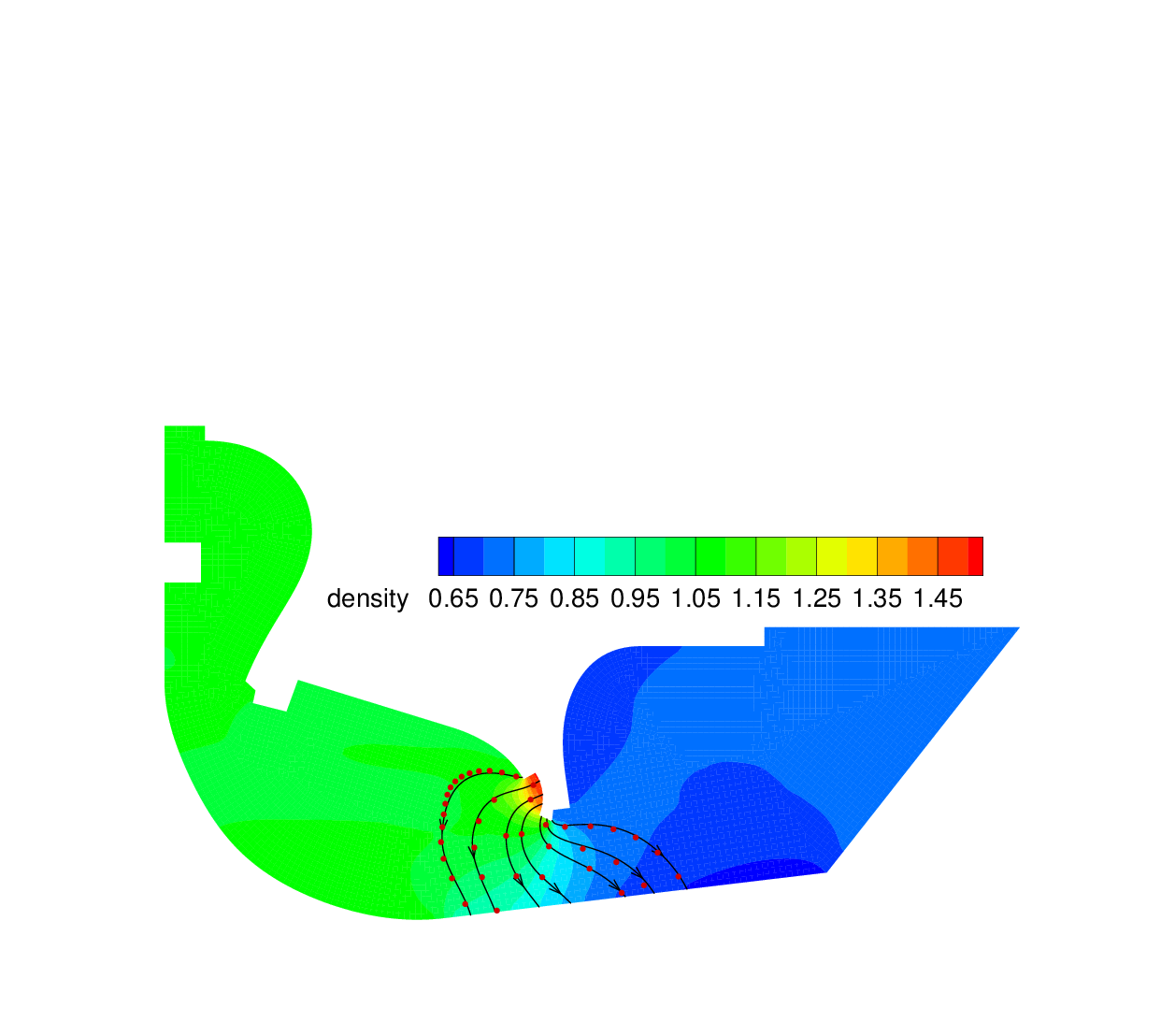}}
    \caption{ 
    (a) The projected mesh in the middle section at $z=0$.
    (b) Velocity mesh independence test under the Cartesian coordinate system and Kn = 0.0426.
    (c) and (d) Velocity mesh independence test under the cylindrical coordinate system, when Kn = 0.0426 and 4.26, respectively.
    }
    \label{mesh_independence}
\end{figure}

Due to the complexity of the divertor's structure, a combination of structured and unstructured grids is employed, with 95\% of the grids being hexahedral. At the junctions of these hexahedral grids, some tetrahedral grids are used. There are 6,268 grids in the middle section, with each grid having an area of approximately about $1~\text{cm}^2$, see Fig.~\ref{mesh_independence}(a).
We run the numerical simulations with 240,000 and 360,000 physical mesh elements,
and find that the two results overlap. This means the grid convergence is achieved and subsequent calculations employ 240,000 physical mesh elements.

We test the velocity mesh independence in the Cartesian coordinates, where the lateral walls in Fig.~\ref{mesh}(b) are assumed to be solid walls with the diffuse gas-surface interaction. Firstly, we use the Newton-Cotes quadrature with $40 \times 40 \times 40$ discrete velocity points and $30 \times 30 \times 30$ discrete velocity points, and find both results overlap. These solutions are used as reference, see the black lines in Fig.~\ref{mesh_independence}(b). 
Then, we gradually replace the Newton-Cotes quadrature by the Gauss-Hermite quadrature, with reduced number of discretized velocities. Eventually we find that applying 10-point Gauss-Hermite quadrature in each velocity direction is adequate the capture the flow field, see red lines with points in Fig.~\ref{mesh_independence}(b). 
The computer resources are summarized in Table~\ref{bimian}. It is seen that, i) the GSIS can find the steady-state solution within 100 iterations, and ii) the simulation time (core hours=CPU cores $\times$ wall-clock time) can be reduced by about two orders of magnitude when the Gauss-Hermite quadrature is applied.


We also test the velocity mesh independence in the cylindrical coordinates, where the lateral walls follow the symmetry boundary condition. 
When Kn=0.0426, we employ 36-point Newton-Cotes quadrature in the $\theta$ direction and 10-point Gauss-Hermite quadrature in $\xi_y$. We also apply the 40-point Newton-Cotes quadrature in the $\xi_r$ direction to obtain a reference solution, see the black lines in Fig.~\ref{mesh_independence}(c). Then, we replace the Newton-Cotes quadrature by the Gauss-Laguerre quadrature in the $\xi_r$ direction, and find that 18-point Gauss-Laguerre quadrature in $\xi_r$ yields results (red dotted lines) identical to the reference solution.
When Kn=4.26, we establish a reference solution using 20-point Gauss-Hermite quadrature in $\xi_y$ and 36-point Gauss-Laguerre quadrature in $\xi_r$, see the black lines in Fig.~\ref{mesh_independence}(d). We observe that identical results (red dotted lines) can also be obtained when using $36\times10\times18$ discrete velocity points.
In the numerical simulation, the GSIS needs 128 cores, 88 iteration steps, and total 44 core hours to find the solution.

\begin{table}[t]
	\centering
	\begin{tabular}{c c c c c c}
		\toprule
	discrete velocity   & CPU cores & times (s) & iteration step & core hours \\
	\midrule
	NC40$\times$NC40$\times$NC40& 640 & 1996 & 94 & 354 \\
	NC30$\times$NC30$\times$NC30 & 640 & 1308 & 92 & 232 \\
	NC30$\times$NC30$\times$GH10 & 320 & 695 & 92 & 61 \\
	NC30$\times$GH10$\times$GH10 & 256 & 199 & 98 & 14 \\
	GH10$\times$GH10$\times$GH10 & 128 & 118 & 93 & 4 \\
		\bottomrule
	\end{tabular}
    \caption{Tests of  velocity mesh independence in the Cartesian coordinates. NC30$\times$GH30$\times$GH10 means that the Newton-Cotes quadrature with 30 uniform grids are used in the $\xi_x$ direction, 30 Gauss-Hermite nodes are used in the $\xi_y$ direction, and 10 Gauss-Hermite nodes are used in the $\xi_z$ direction.
    The Knudsen number is 0.0426, and the convergence criterion is the relative difference in density and temperature between two consecutive iterations  is less than $10^{-5}$. 
    }
    \label{bimian}
\end{table}
\par



Compare with the DSMC simulation~\cite{tantos2023dtt}, where 40 million spatial cells and 0.688 million CPU core hours are used, the GSIS can be faster by about 4 orders of magnitude\footnote{
Note that we employ the cylindrical coordinates to precisely implement the symmetry boundary conditions at the two lateral walls, which is less efficient compared to Cartesian coordinates. For example, if the full divertor geometry (360°) were simulated using Cartesian coordinates instead of the 20° section shown in Fig.~\ref{mesh}, with the $10 \times 10 \times 10$ Gauss-Hermite quadrature, the core hours required would be $4 \times 18 = 72$. In contrast, the DSMC method would require $18\times0.688$ million core hours, which is slower by five orders of magnitude. Simulating the full-scale divertor is advantageous, however, since in reality, perfect toroidal symmetry is not always achieved.
}. This huge reduction of computational time is due to i) the asymptotic-preserving property of the GSIS, which allows use of coarse spatial grids, ii) the fast convergence property of the GSIS, which finds the steady state solution within 100 iterations, and iii) the deterministic nature of the GSIS, which does not suffer from the statistic error. It should be noted that the magnetic coils and the cooling pipes are not considered in the simplified geometry in Fig.~\ref{mesh}(b,c). Even if they are added, the number of the spatial grids in GSIS would at most increase by 10 times, and the GSIS can still be faster than the DSMC by 3 orders of magnitude.



\begin{figure}[p]
    \centering
    {\includegraphics[trim={100 60 100 100},clip,width=0.3\linewidth]{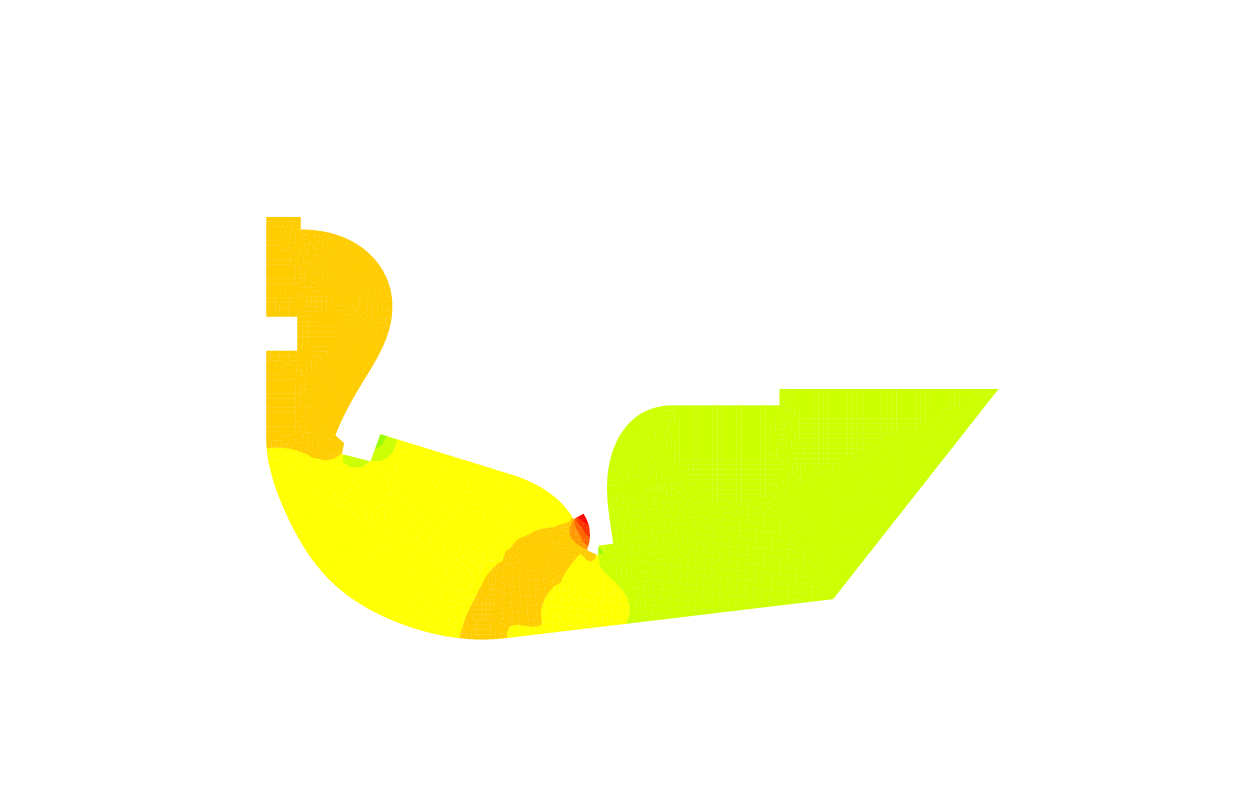}}
    {\includegraphics[trim={100 60 100 100},clip,width=0.3\linewidth]{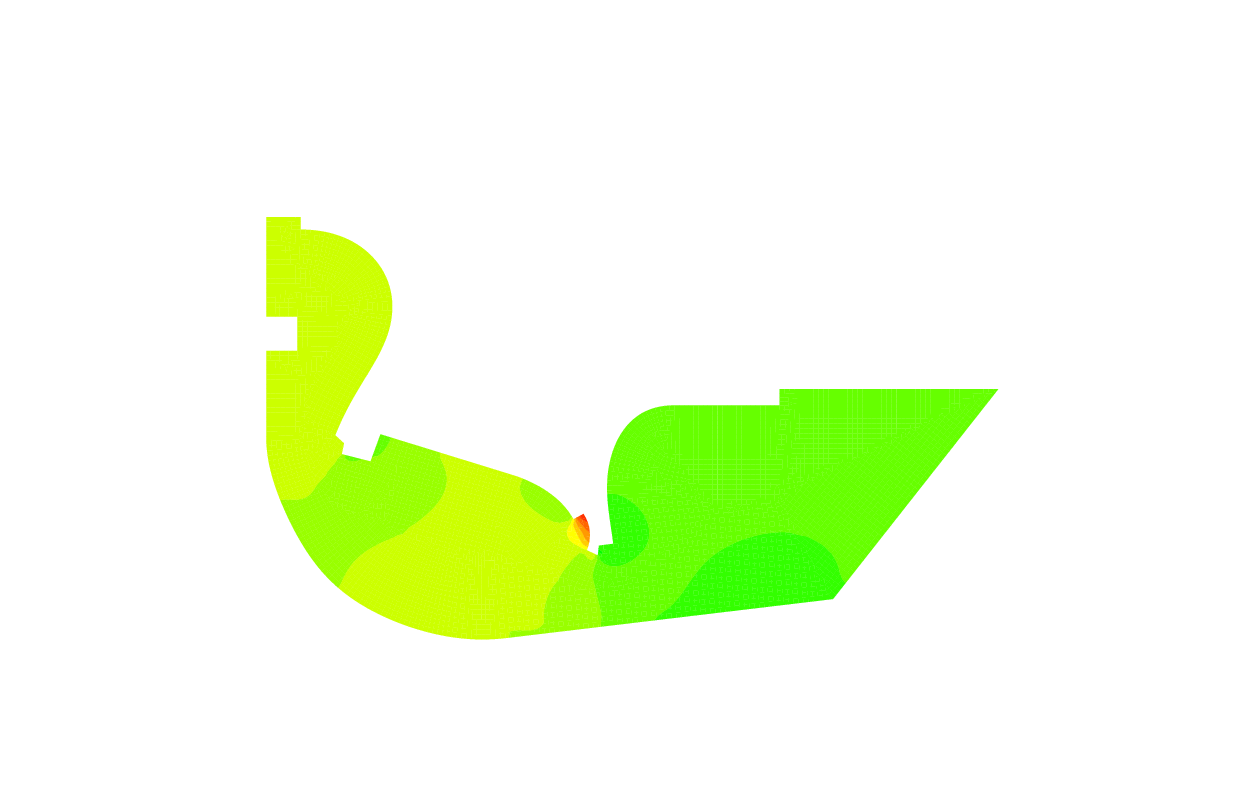}}
    {\includegraphics[trim={100 60 20 100},clip,width=0.35\linewidth]{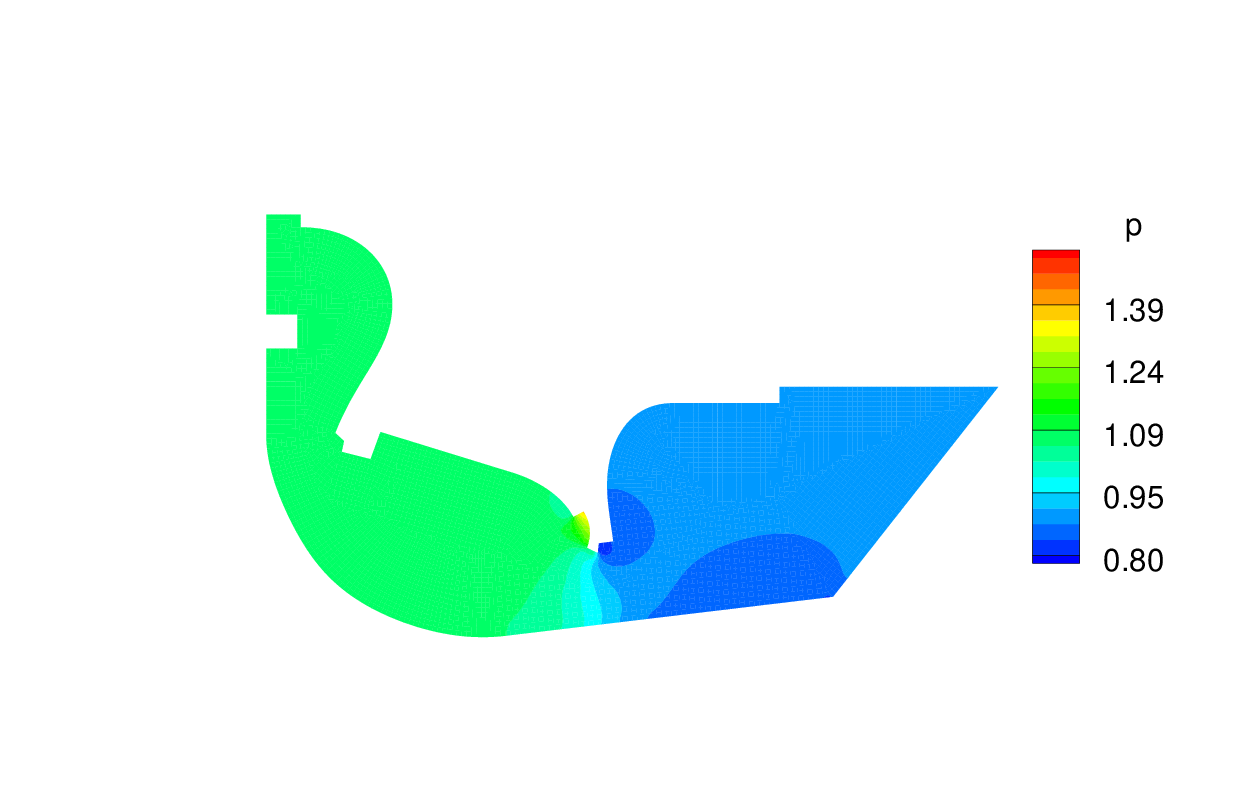}}\\
    {\includegraphics[trim={100 60 100 100},clip,width=0.3\linewidth]{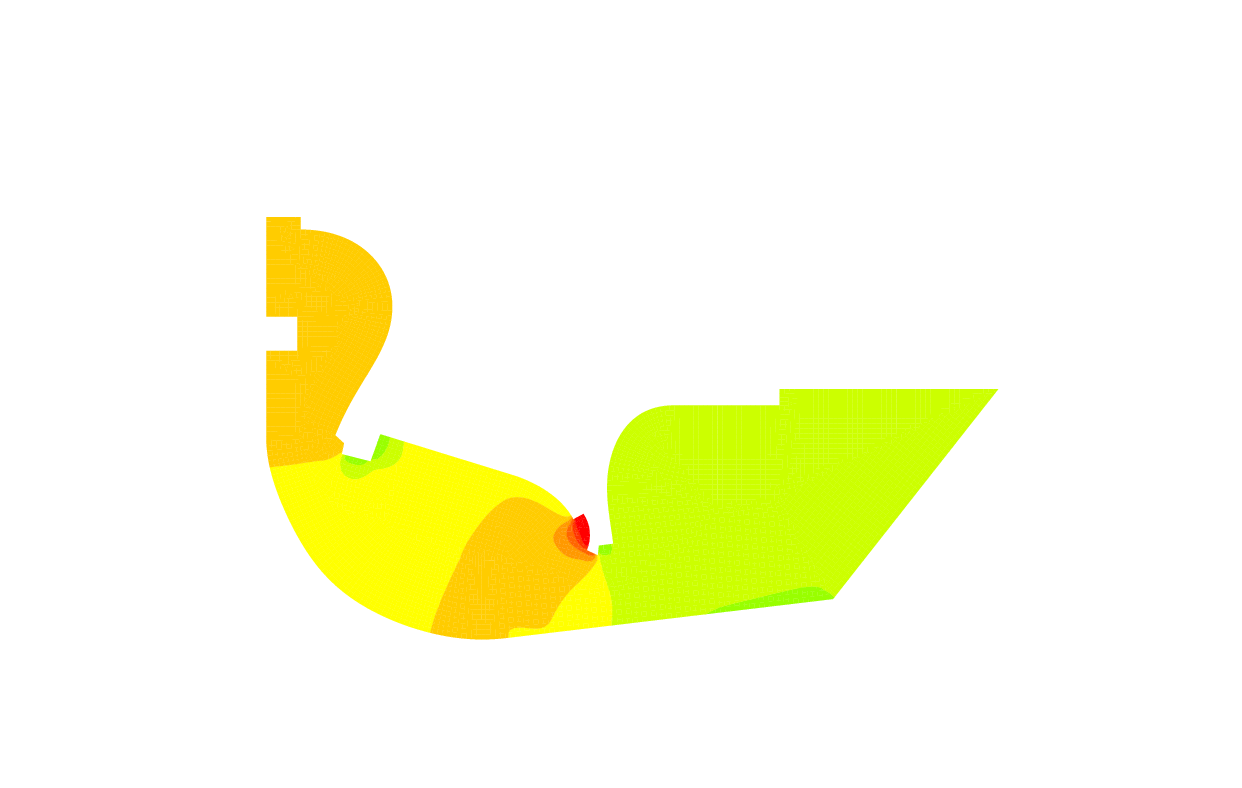}}
    {\includegraphics[trim={100 60 100 100},clip,width=0.3\linewidth]{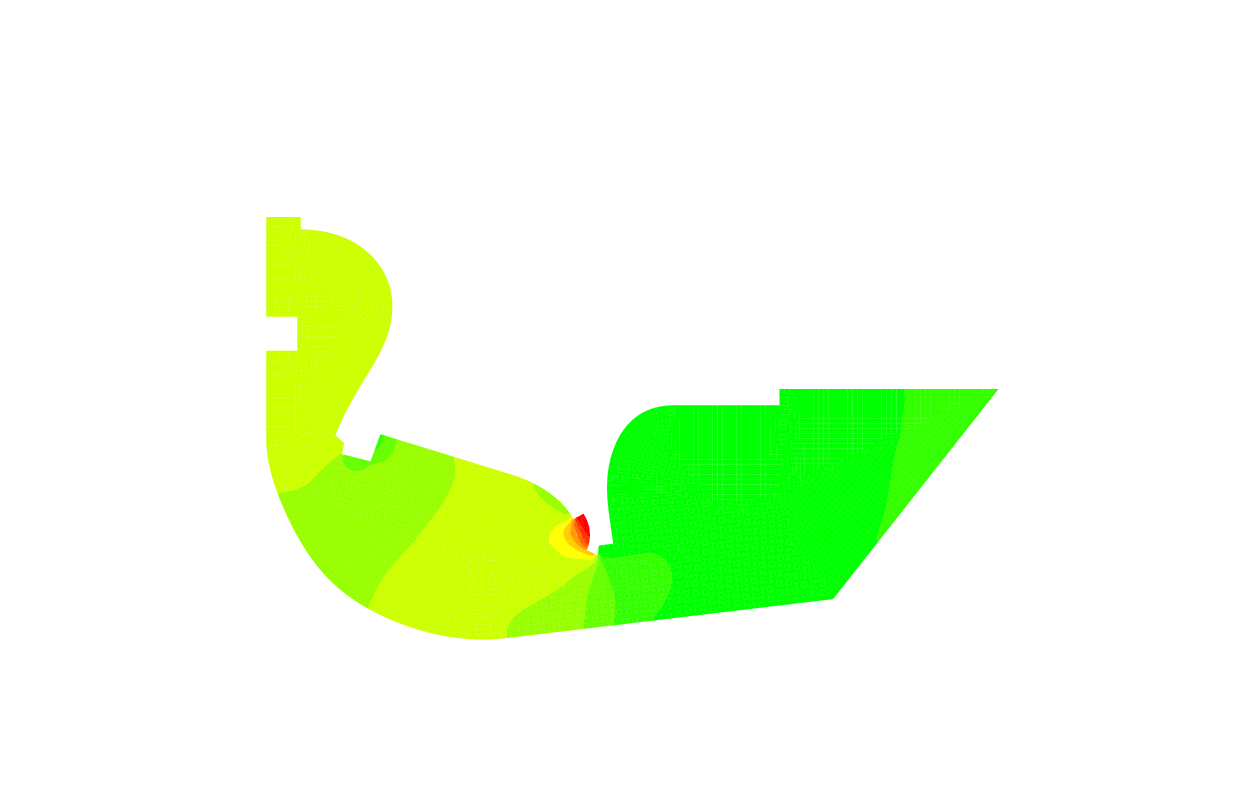}}
    {\includegraphics[trim={100 60 20 100},clip,width=0.35\linewidth]{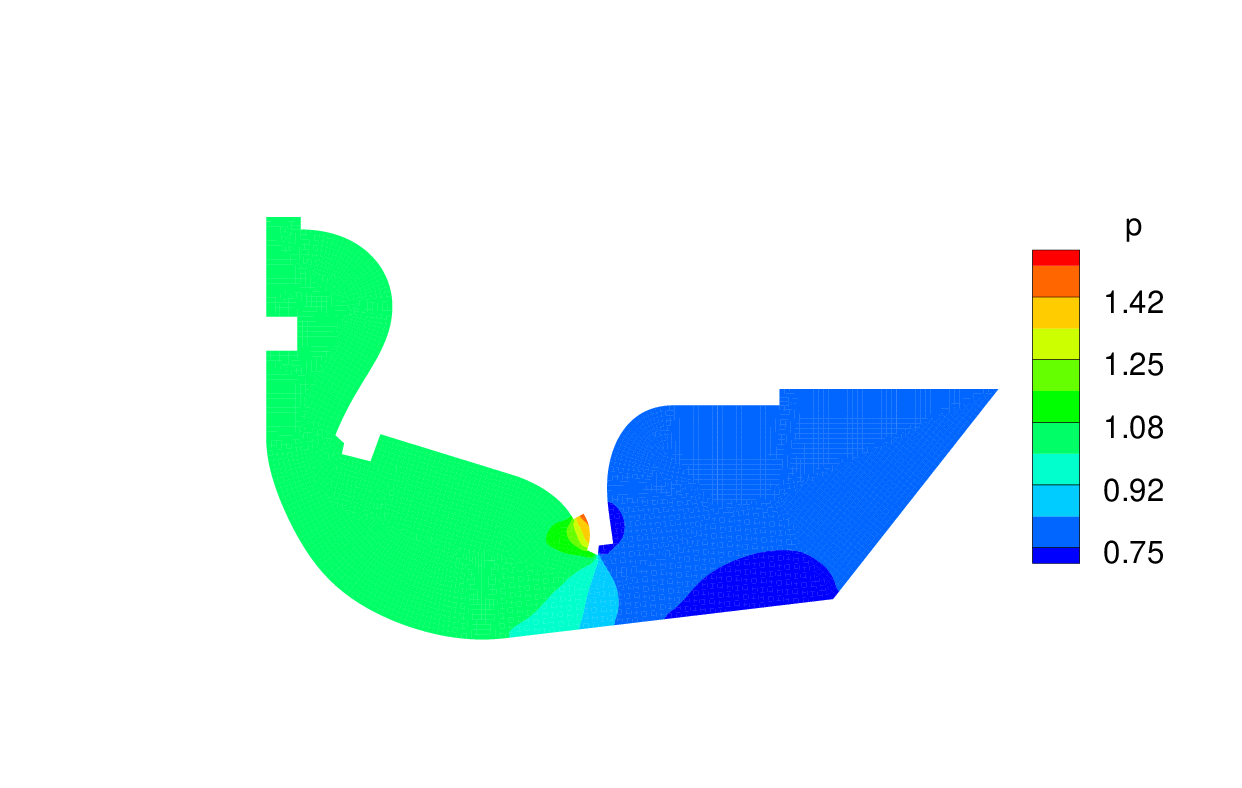}}\\
    {\includegraphics[trim={100 60 100 100},clip,width=0.3\linewidth]{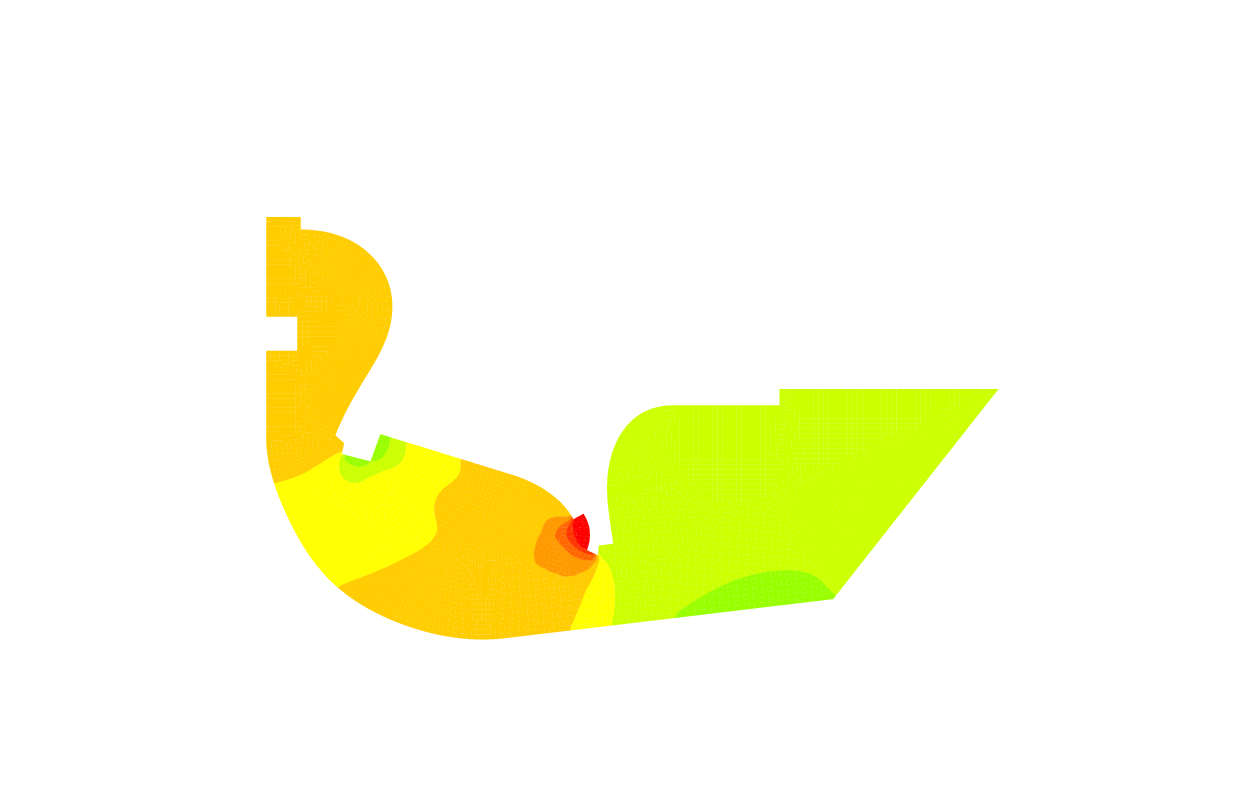}}
    {\includegraphics[trim={100 60 100 100},clip,width=0.3\linewidth]{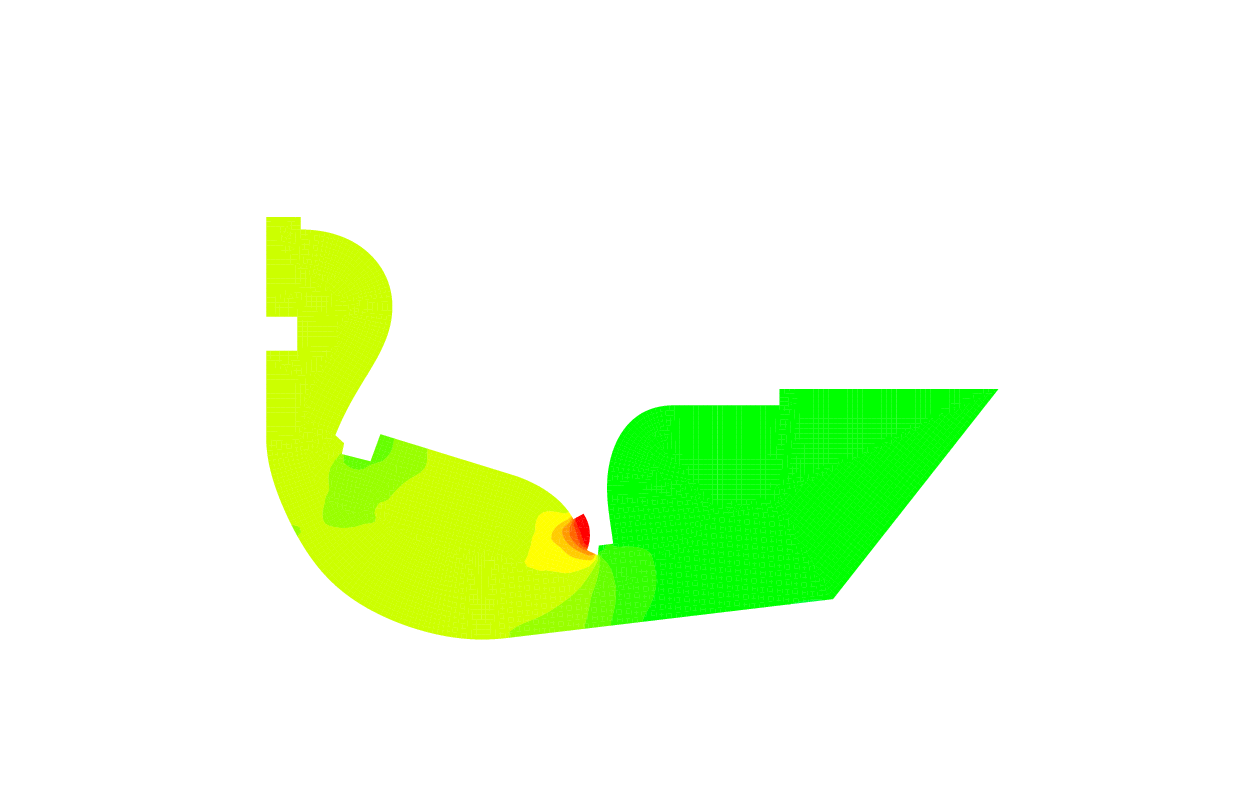}}
    {\includegraphics[trim={100 60 20 100},clip,width=0.35\linewidth]{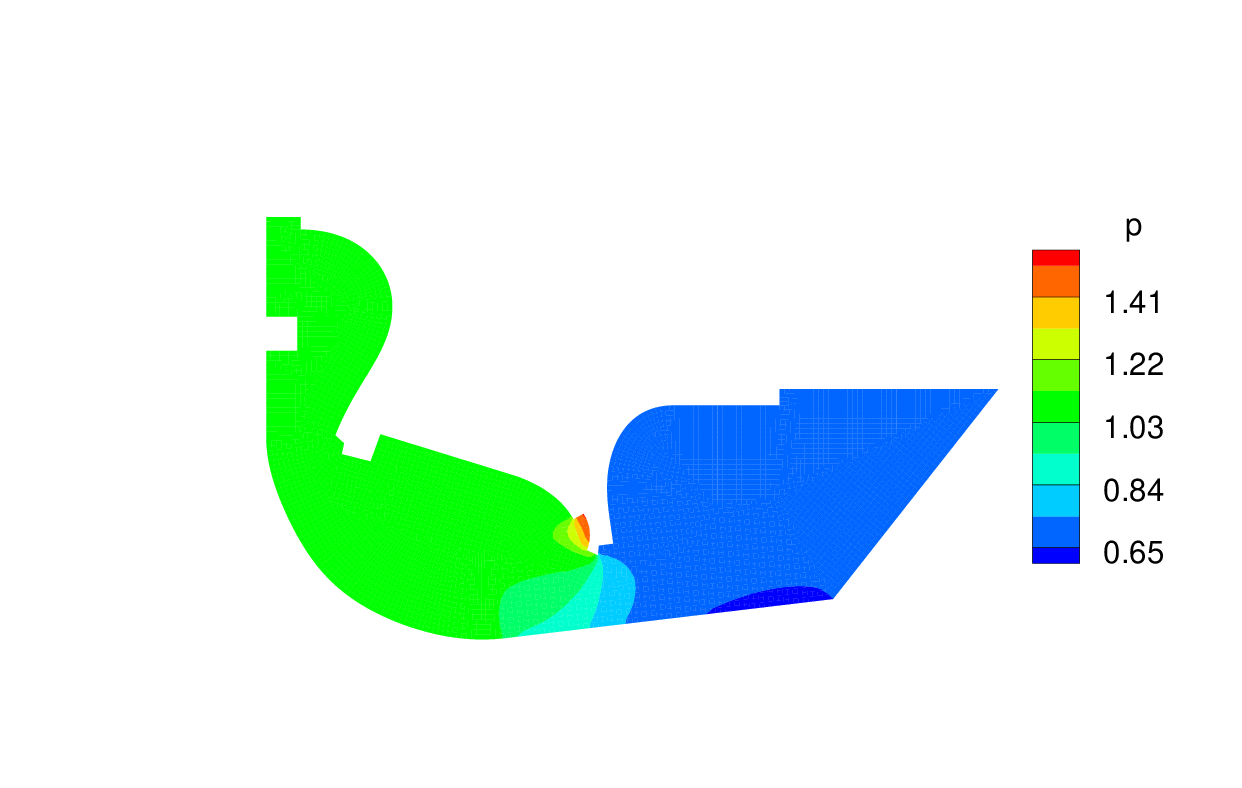}}
    \caption{Pressure distributions at the midsection ($z=0$). The absorptivity from the left column to right column is 0.05, 0.1, 0.3, respectively. The Knudsen number from the top row to bottom row is 0.0426, 0.426, 4.26, respectively.
    }
    \label{p426}
\end{figure}

\begin{figure}[p]
    \centering
    {\includegraphics[trim={50 80 100 10},clip,width=0.3\linewidth]{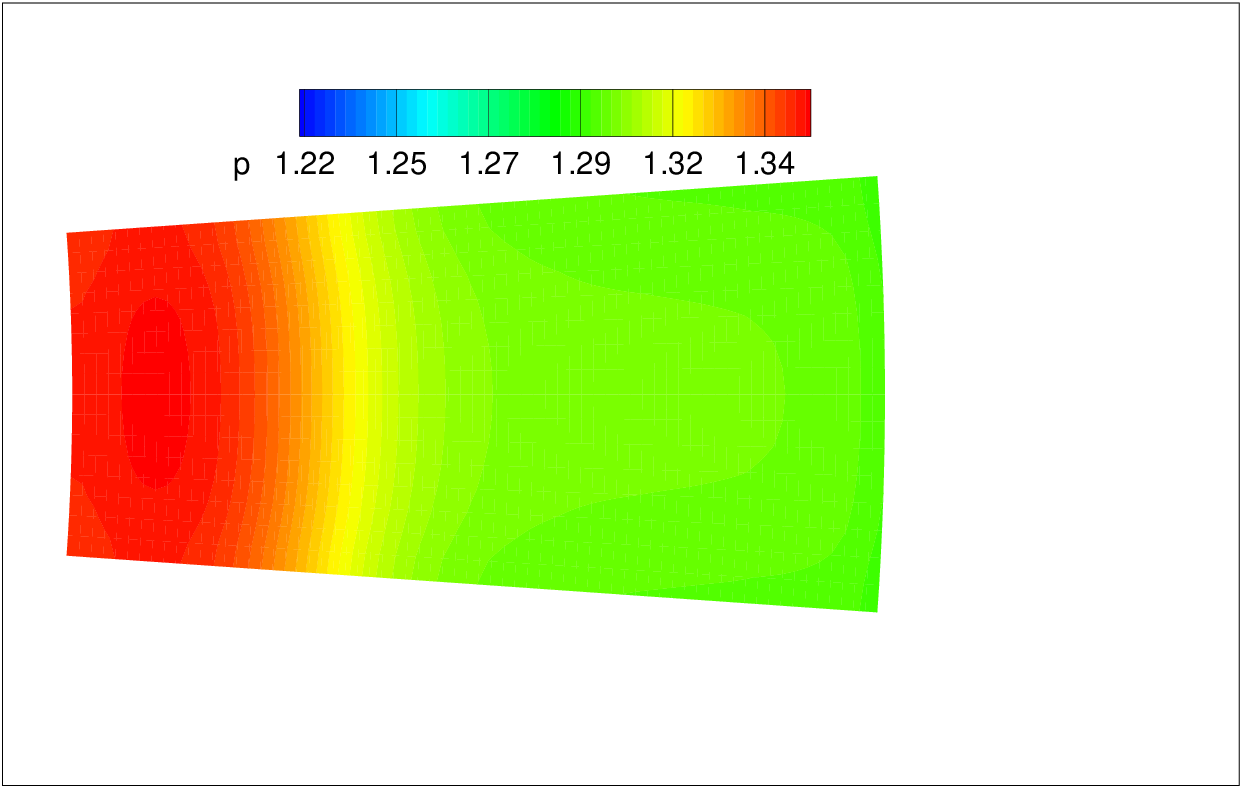}}
    {\includegraphics[trim={50 80 100 10},clip,width=0.3\linewidth]{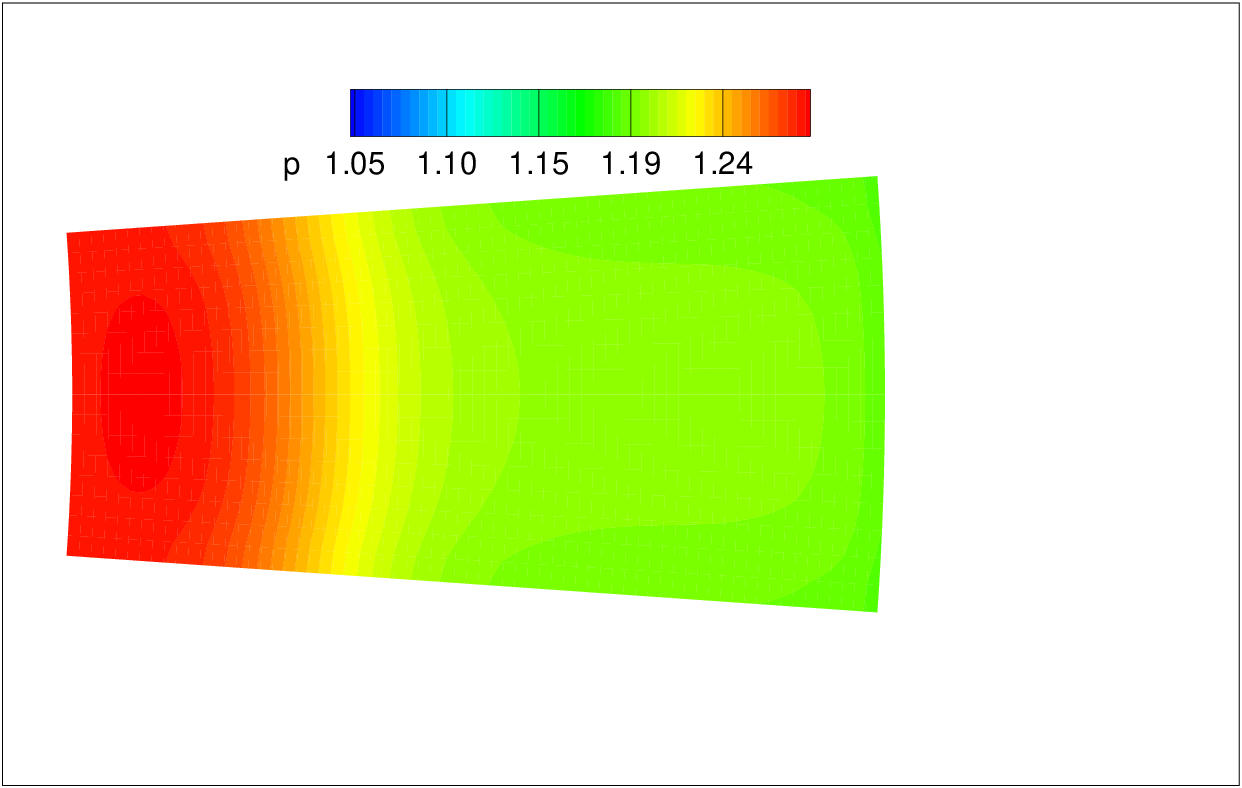}}
    {\includegraphics[trim={50 80 100 10},clip,width=0.3\linewidth]{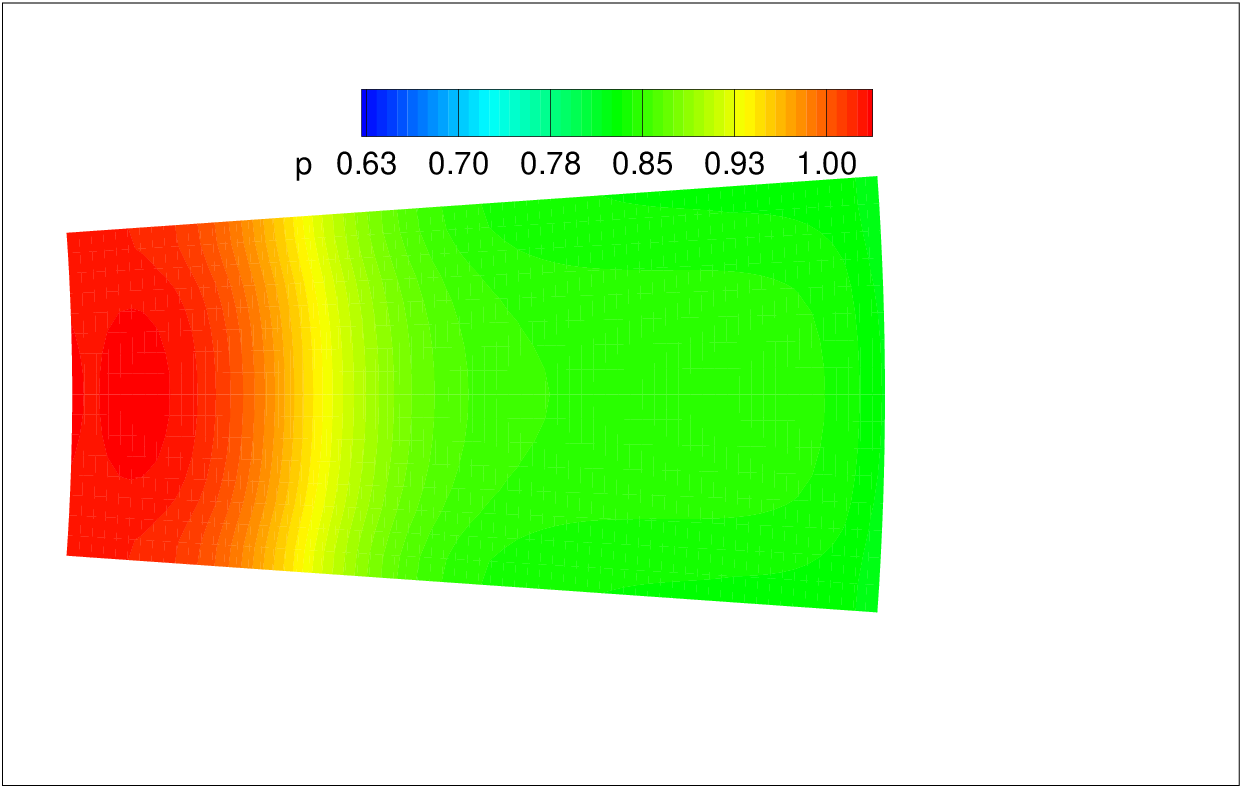}}\\
    {\includegraphics[trim={50 80 100 50},clip,width=0.3\linewidth]{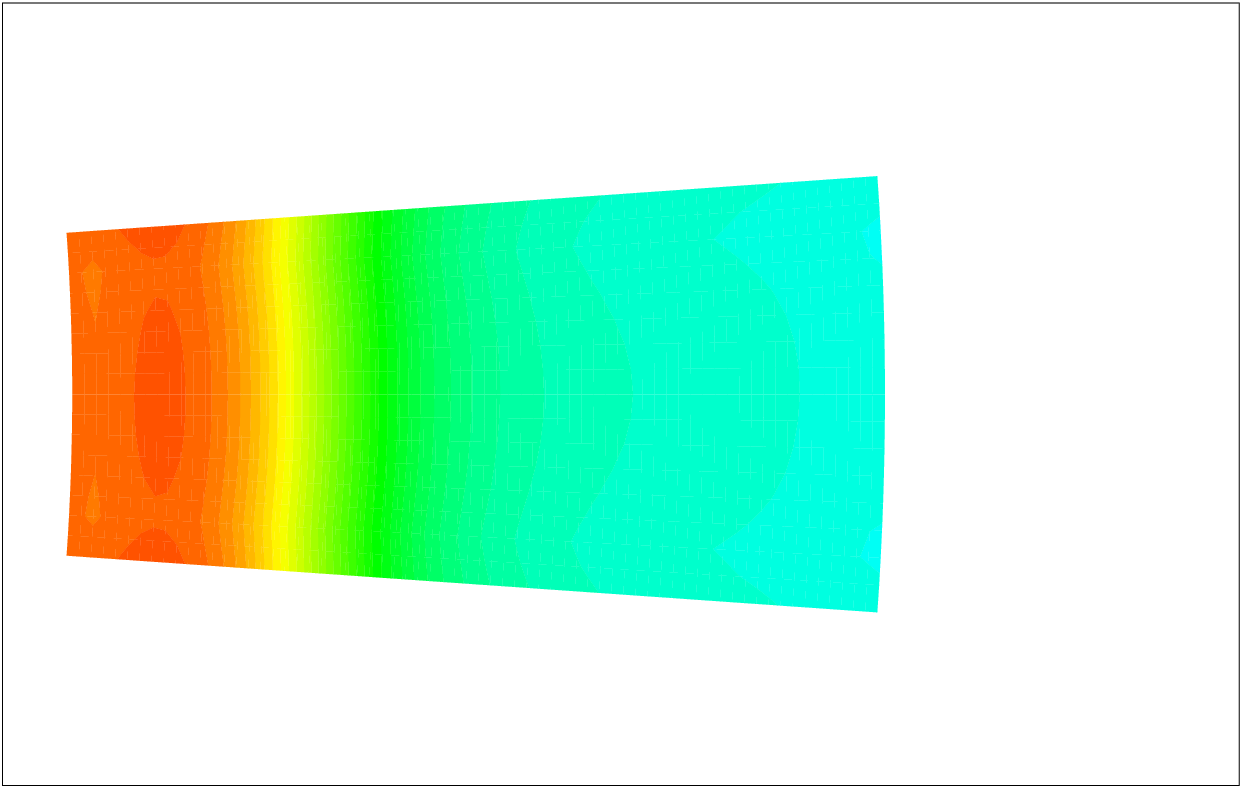}}
    {\includegraphics[trim={50 80 100 50},clip,width=0.3\linewidth]{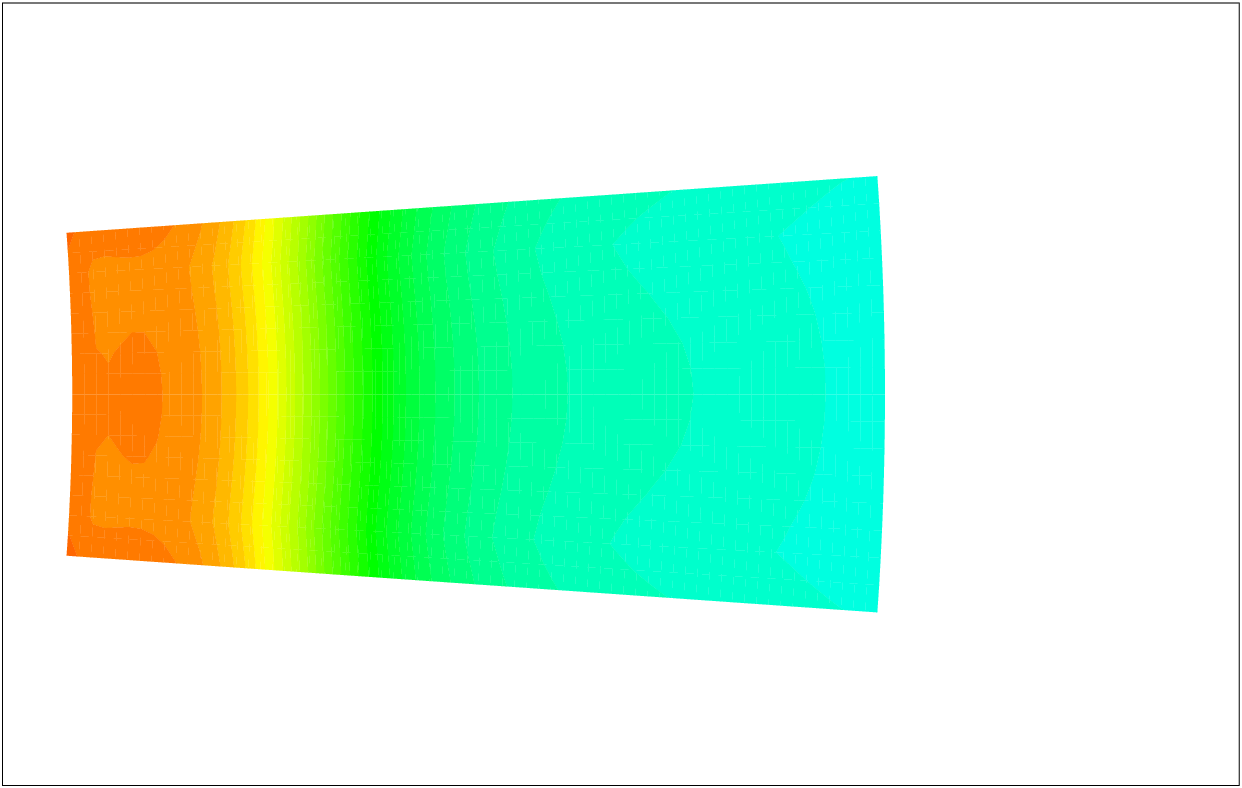}}
    {\includegraphics[trim={50 80 100 50},clip,width=0.3\linewidth]{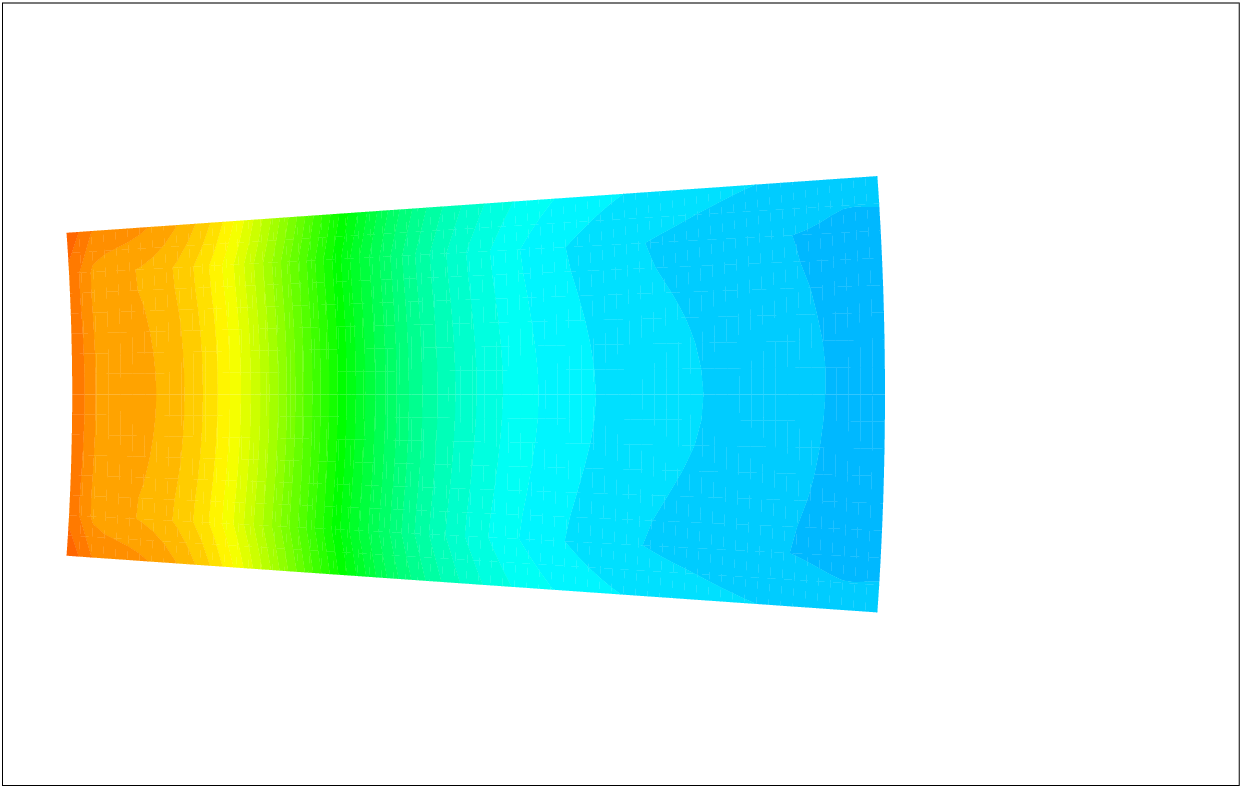}}\\
    {\includegraphics[trim={50 80 100 50},clip,width=0.3\linewidth]{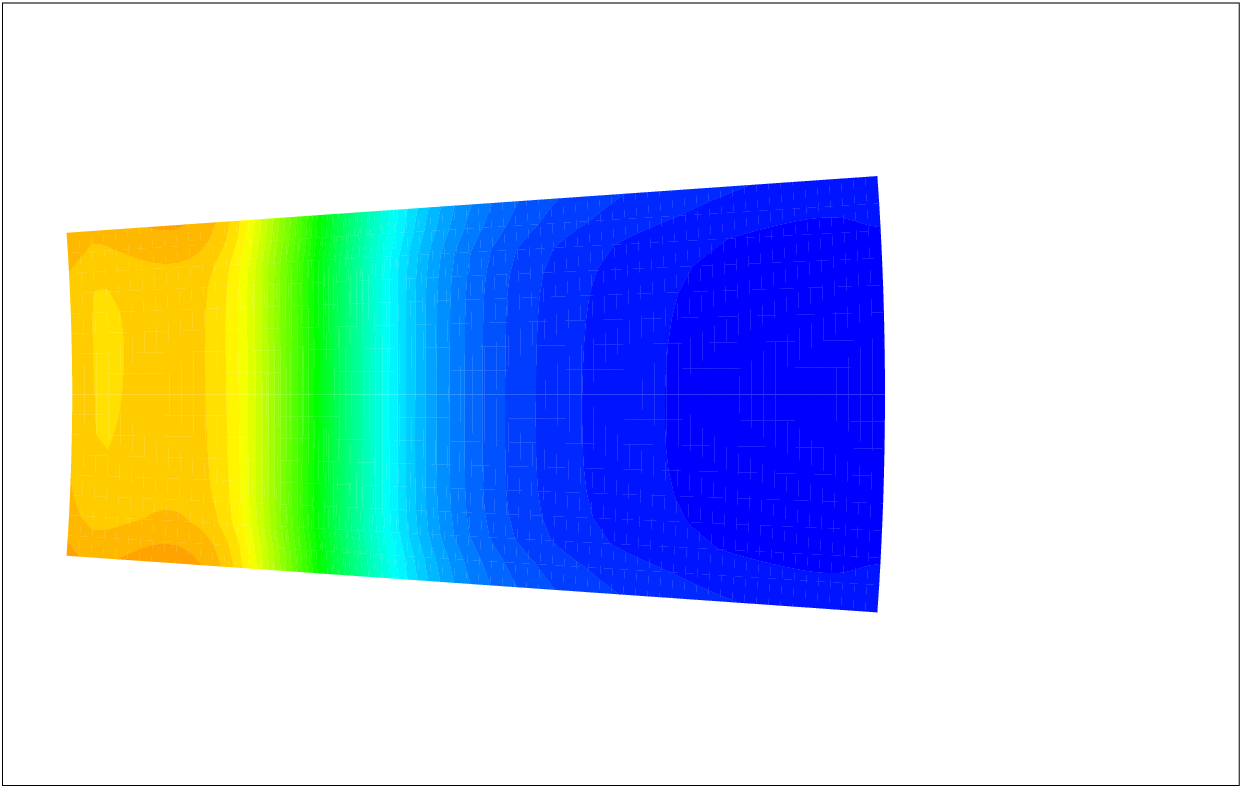}}
    {\includegraphics[trim={50 80 100 50},clip,width=0.3\linewidth]{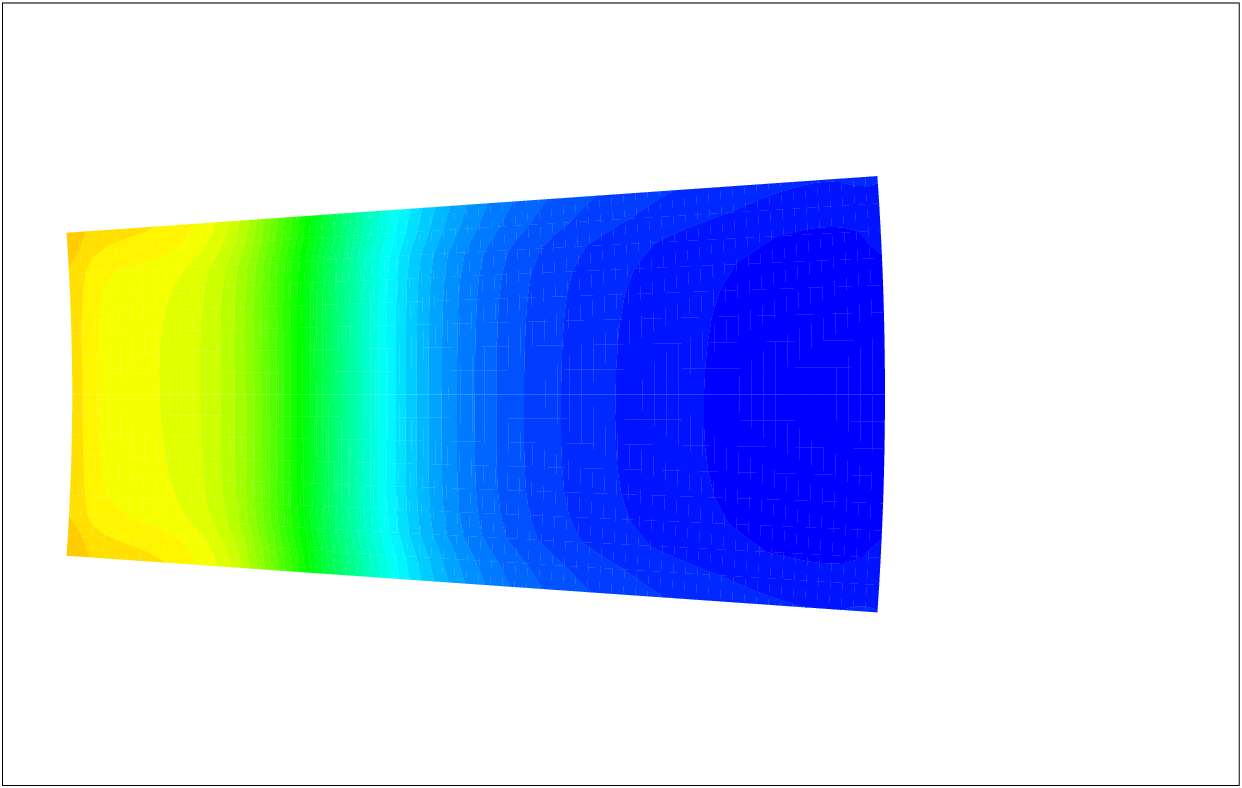}}
    {\includegraphics[trim={50 80 100 50},clip,width=0.3\linewidth]{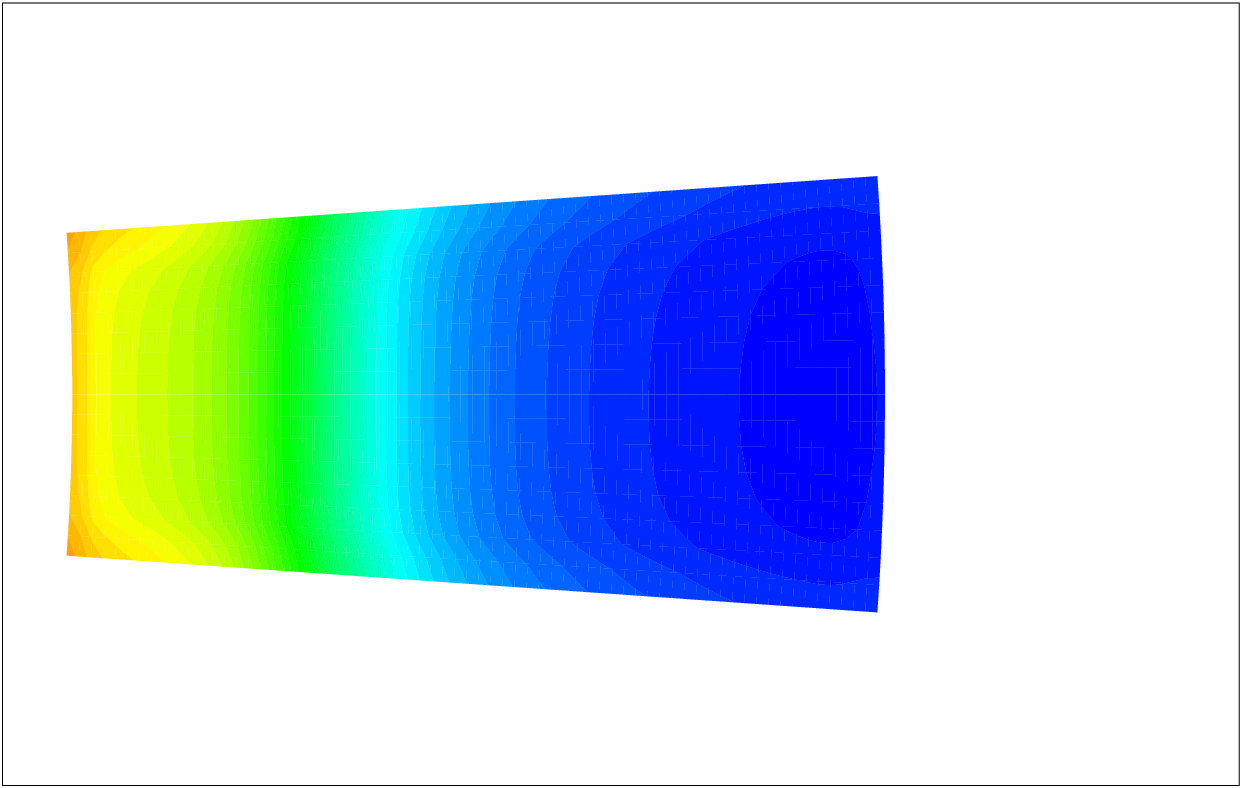}}
    \caption{Pressure distributions at the absorption pump, which is located in the middle of the bottom of the divertor, see the blue area in Fig.~\ref{mesh}(b,c). The absorptivity from the left column to right column is 0.05, 0.1, 0.3, respectively. The Knudsen number  from the top row to bottom row is 0.00426, 0.0426, 0.426, respectively.
    }
    \label{pump426}
\end{figure}

\section{Roles of Knudsen numbers, absorptivity and  temperature }

In this section, we employ the GSIS with the cylindrical coordinates in the molecular velocity space to study the flow fields within the divertor and analyze the pumping speed under various absorptivity, temperature, and Knudsen numbers. All results are normalized using the following parameters:
\begin{equation}
    T_{ref} = T_0, \quad 
    v_{ref}==\sqrt{2k_BT_0/m},
    \quad 
    p_{ref} = \rho_0 R T_0.
\end{equation}

\subsection{Role of Knudsen number and absorptivity}

We first consider the case where the temperature of the inlet gas and the surface temperature is $T_0=300$~K.
Figure \ref{p426} shows the pressure distribution in the midsection ($z = 0$) of the divertor at varying Knudsen numbers and absorptivities. Because the flow velocity is much smaller than the sound of speed, the gas flow exhibits minimal temperature fluctuations inside the divertor. Hence, the pressure can be deemed directly proportional to density. As the Knudsen number rises, the pressure variation within the divertor gradually increases, rendering the flow characteristics more intricate.
As the absorptivity rises, both the pressure variation and the extent of the low-pressure zone increase. 
Figure~\ref{pump426} illustrates the pressure distribution at the outlet. With the increase of Knudsen number, the pressure difference increases, and the pressure distribution becomes more complex. As the absorptivity increases, the pressure decreases, which means more gas escapes from the outlet pump. 

\begin{figure}[t]
    \centering
    \subfloat[$\zeta=0.05$]{\includegraphics[trim={0 0 0 50},clip,width=0.4\linewidth]{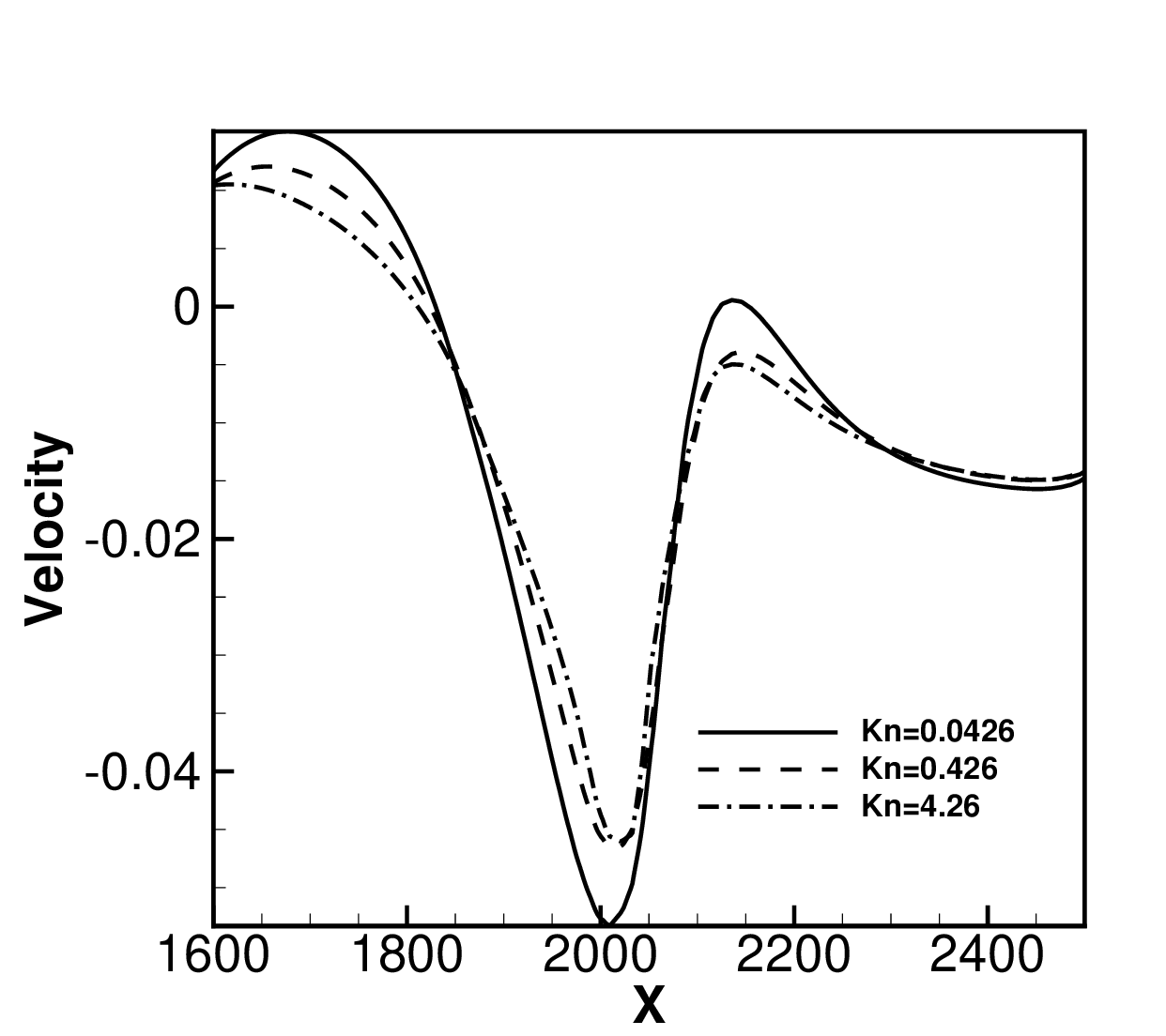}}
    \subfloat[$\zeta=0.1$]{\includegraphics[trim={0 0 0 50},clip,width=0.4\linewidth]{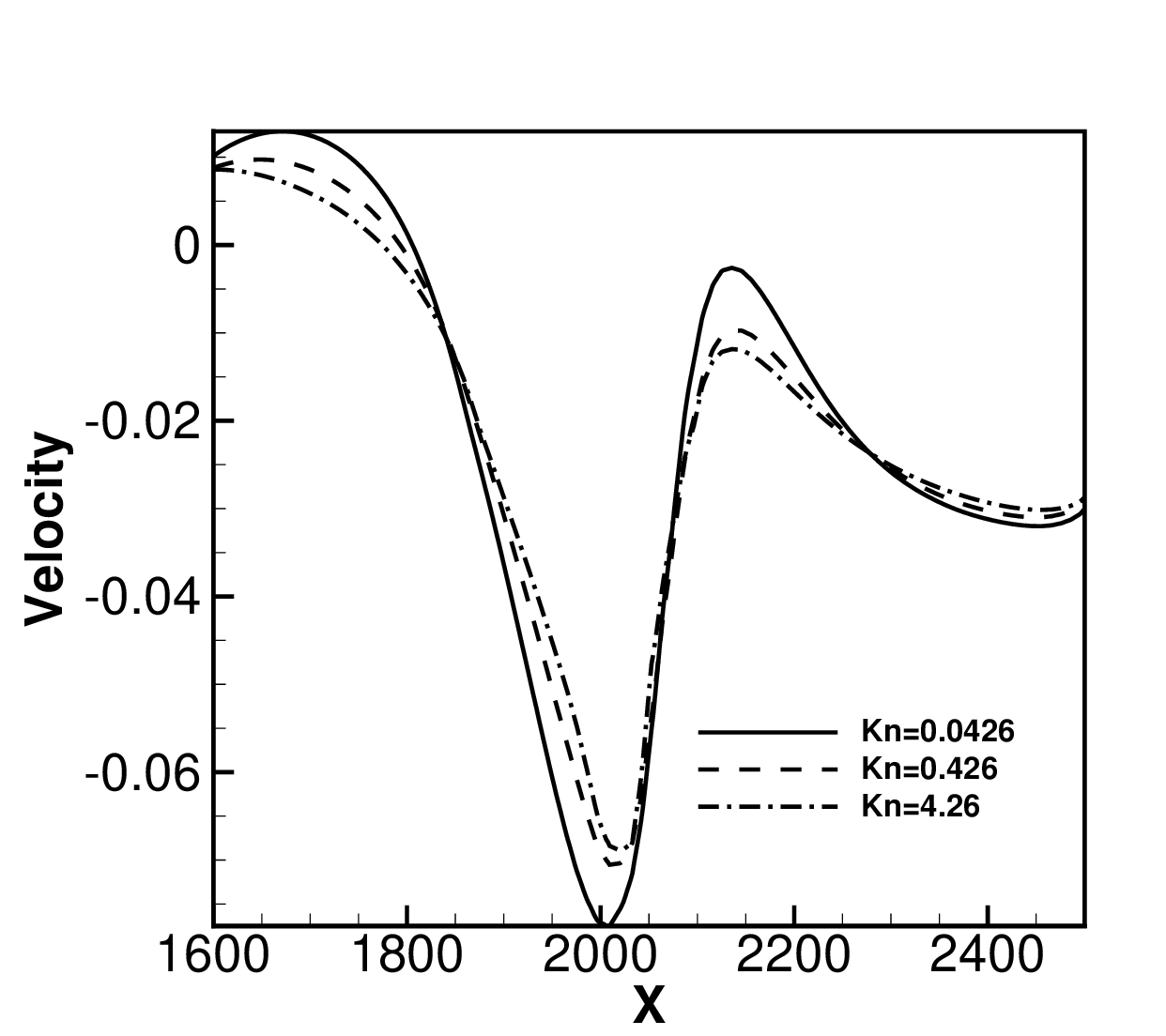}}\\
    \subfloat[$\zeta=0.3$]{\includegraphics[trim={0 0 0 50},clip,width=0.4\linewidth]{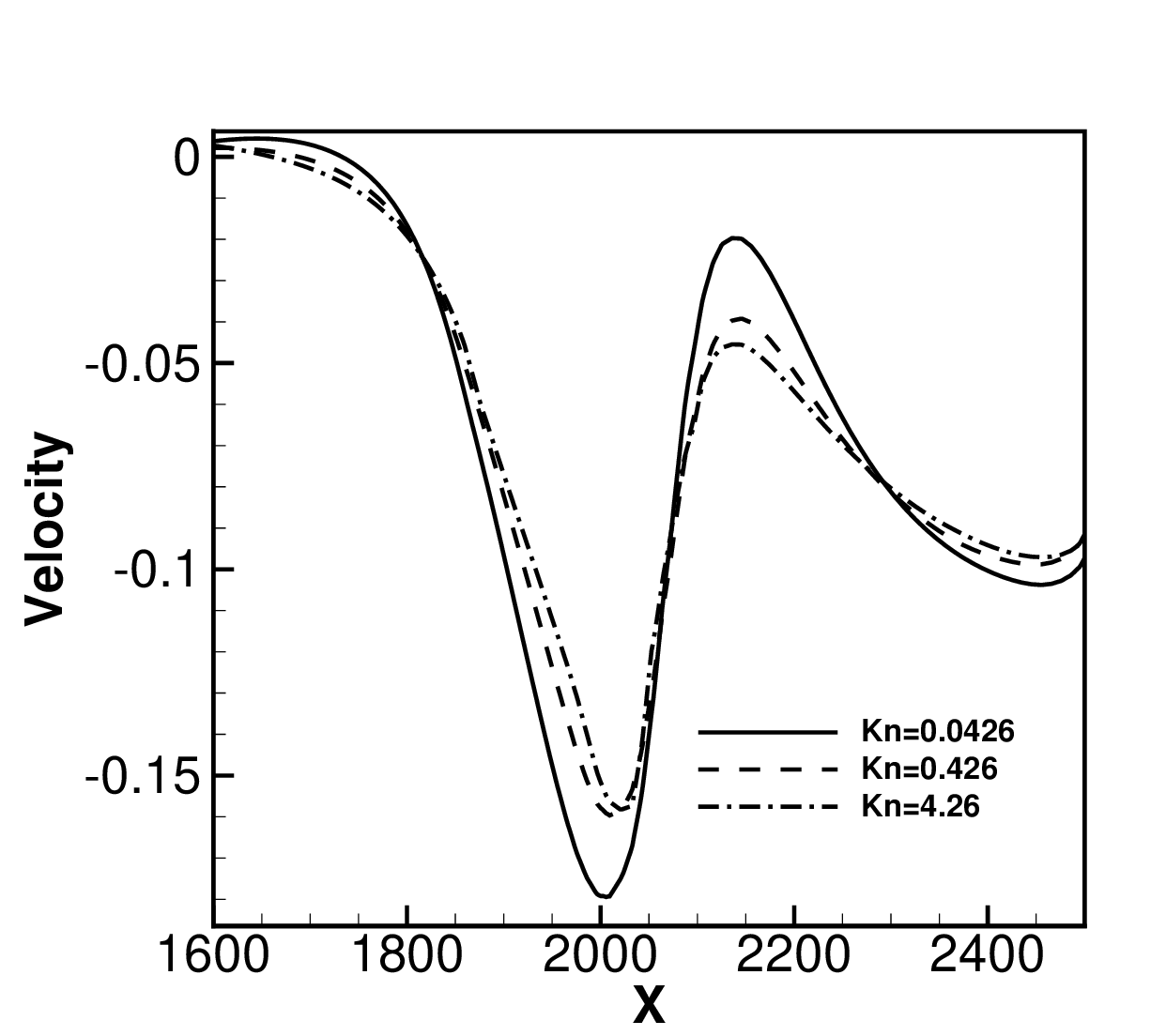}}
     \subfloat[Mass flow rate]{\includegraphics[trim={0 0 0 50},clip,width=0.4\linewidth]{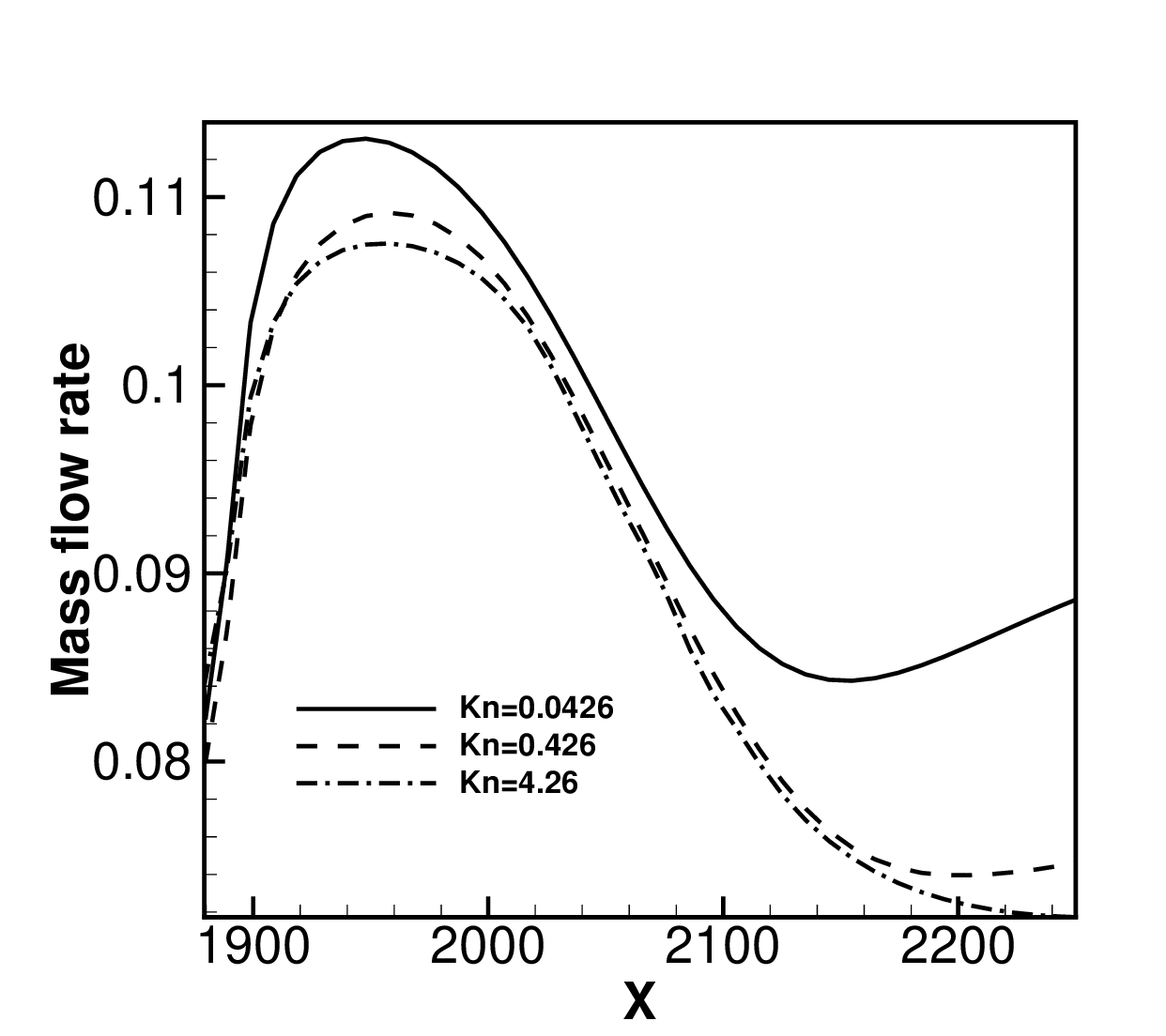}}
    \caption{
    (a-c) The velocity in the y-direction of midsection ($z = 0$). (d) Mass flow distribution along the central axis ($z = 0$) of the absorption pump for different Knudsen numbers (absorptivity = 0.3). Both the gas temperature at the inlet and the wall temperature are 300~K. 
    }
    \label{zhong_velocity}
\end{figure}


The reduced density and pressure at the pump opening implies higher mass flow rate at higher absorptivity.
Figure~\ref{zhong_velocity}(a-c) shows the velocity distribution in the $y$-direction, across the divertor cross-section ($z=0$). The outlet pump is located within the range of $1850 \le x \le 2250$. In this region, as $x$ increases, the gas speed first increases and reaches its maximum near $x=2000$, then it decreases and reaches its minimum value around $x=2150$. Finally it gradually increases. The Knudsen number has small impact on the magnitude of the flow velocity, however, increasing the absorptivity will significantly enhance the velocity, thereby increasing the mass flow rate of the pump.

Figure~\ref{zhong_velocity}(d) shows the distribution of mass flow rate $\rho\bm{u}\cdot\bm{n}$ along the central axis of the absorption pump, assuming a constant absorptivity of 0.3, where $\bm{n}$ represents the outward normal vector at the pump opening. The maximum mass flow rate occurs at approximately $x_m=1950$, which is slightly to the left of the location where the speed is at its peak. For $x<x_m$, the mass flow rate increases monotonically with $x$. Conversely, for $x>x_m$, a monotonic decrease in the mass flow rate is observed only at large Knudsen numbers, such as Kn=4.26. At smaller Knudsen numbers, such as Kn=0.426 and 0.0426, a local minimum in the mass flow rate is present, that is, when $x$ increases, the mass flow rate first decreases and then increases.

\begin{figure}[t]
    \centering
    {\includegraphics[trim={0 0 50 200},clip,width=0.32\linewidth]{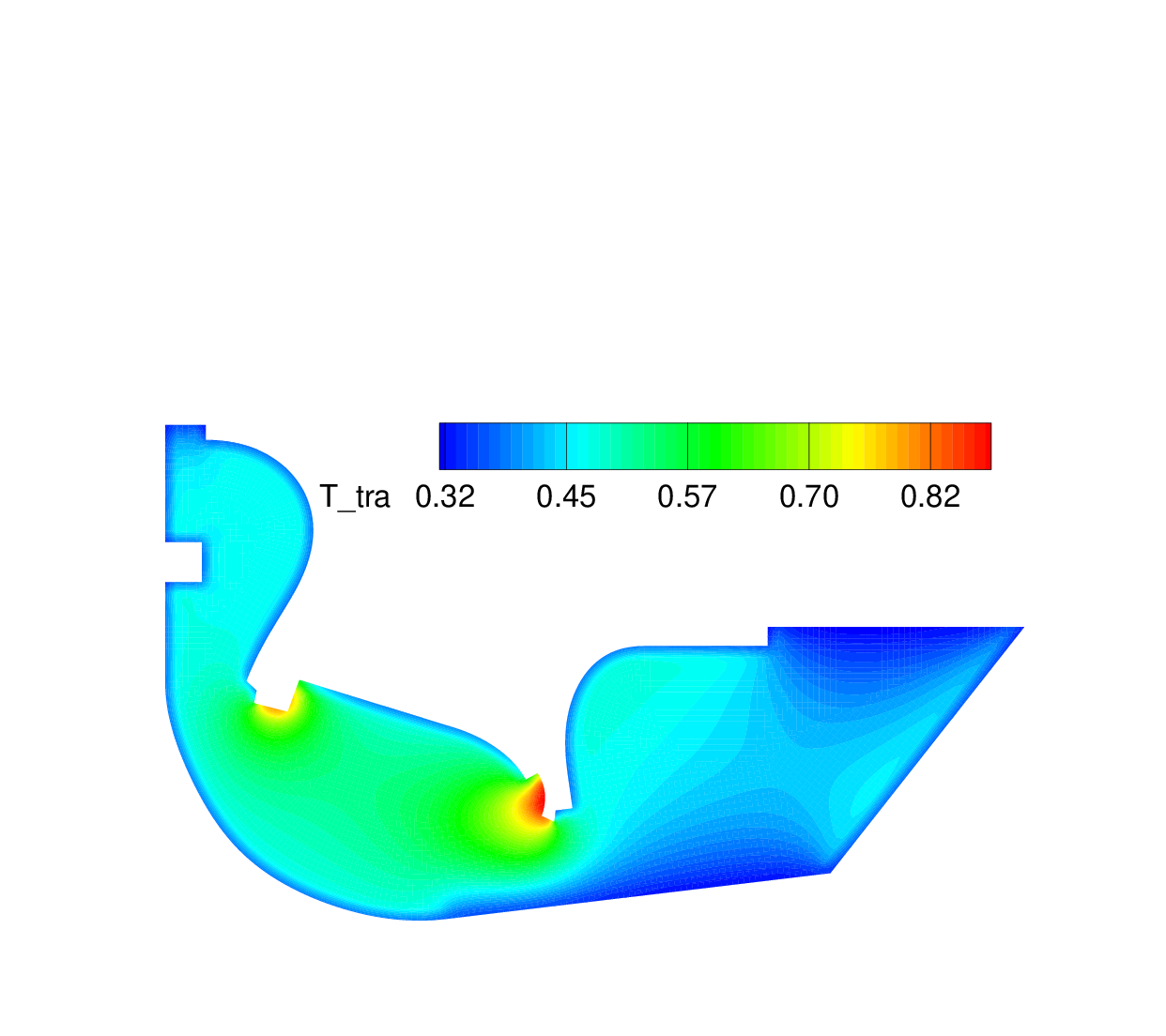}}
    {\includegraphics[trim={0 0 50 200},clip,width=0.32\linewidth]{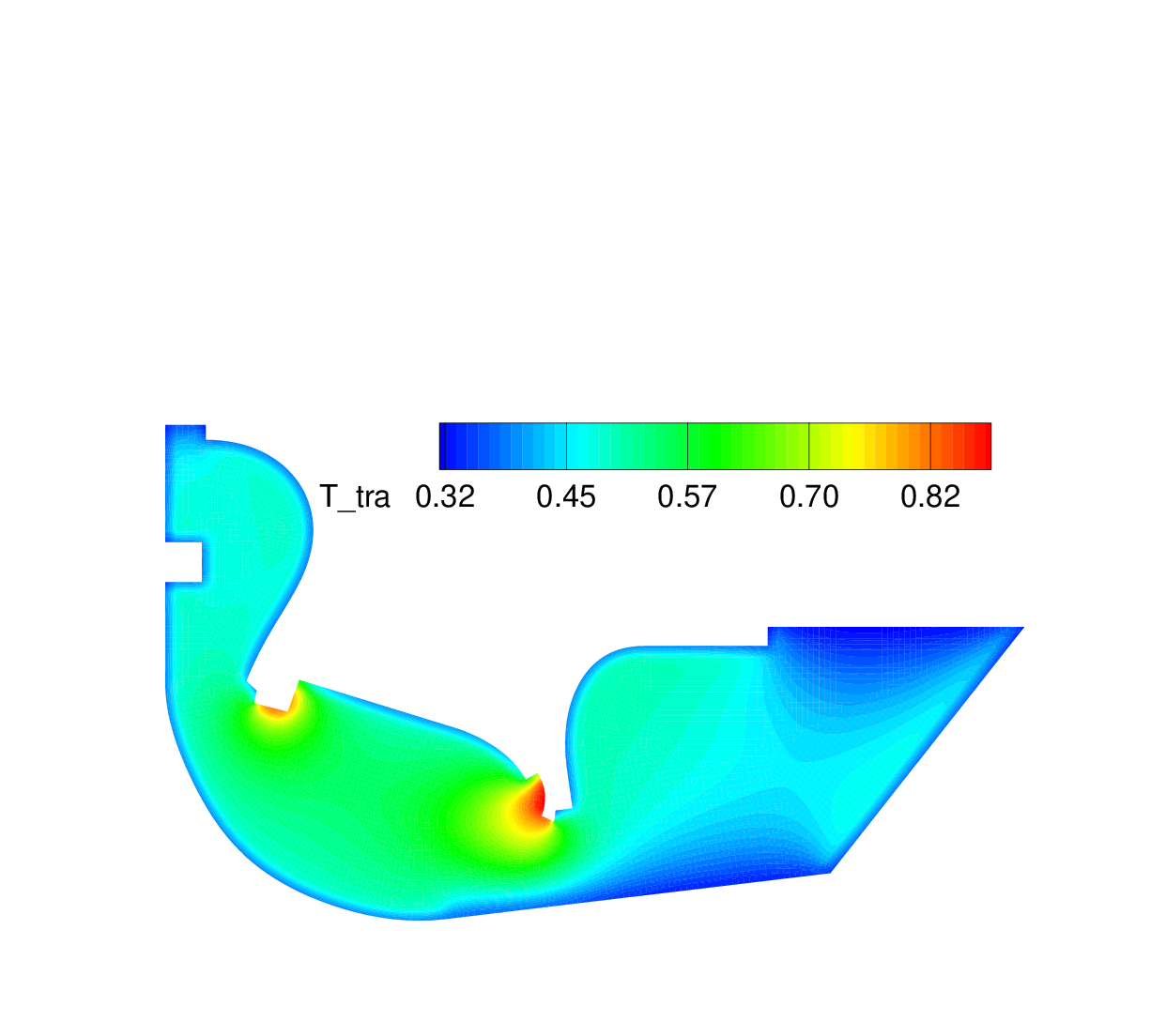}}
    {\includegraphics[trim={0 0 50 200},clip,width=0.32\linewidth]{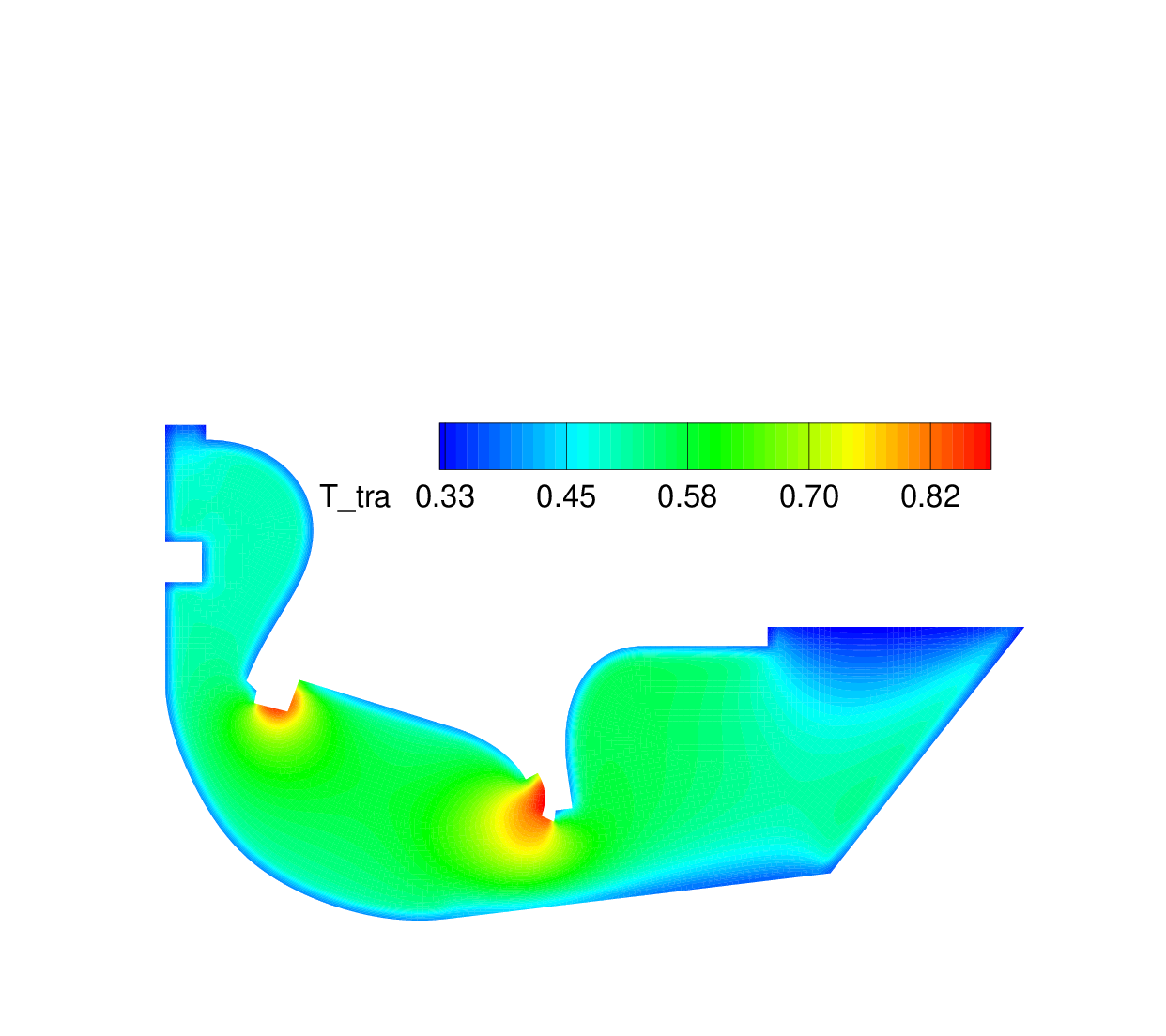}}\\
    {\includegraphics[trim={0 0 50 200},clip,width=0.32\linewidth]{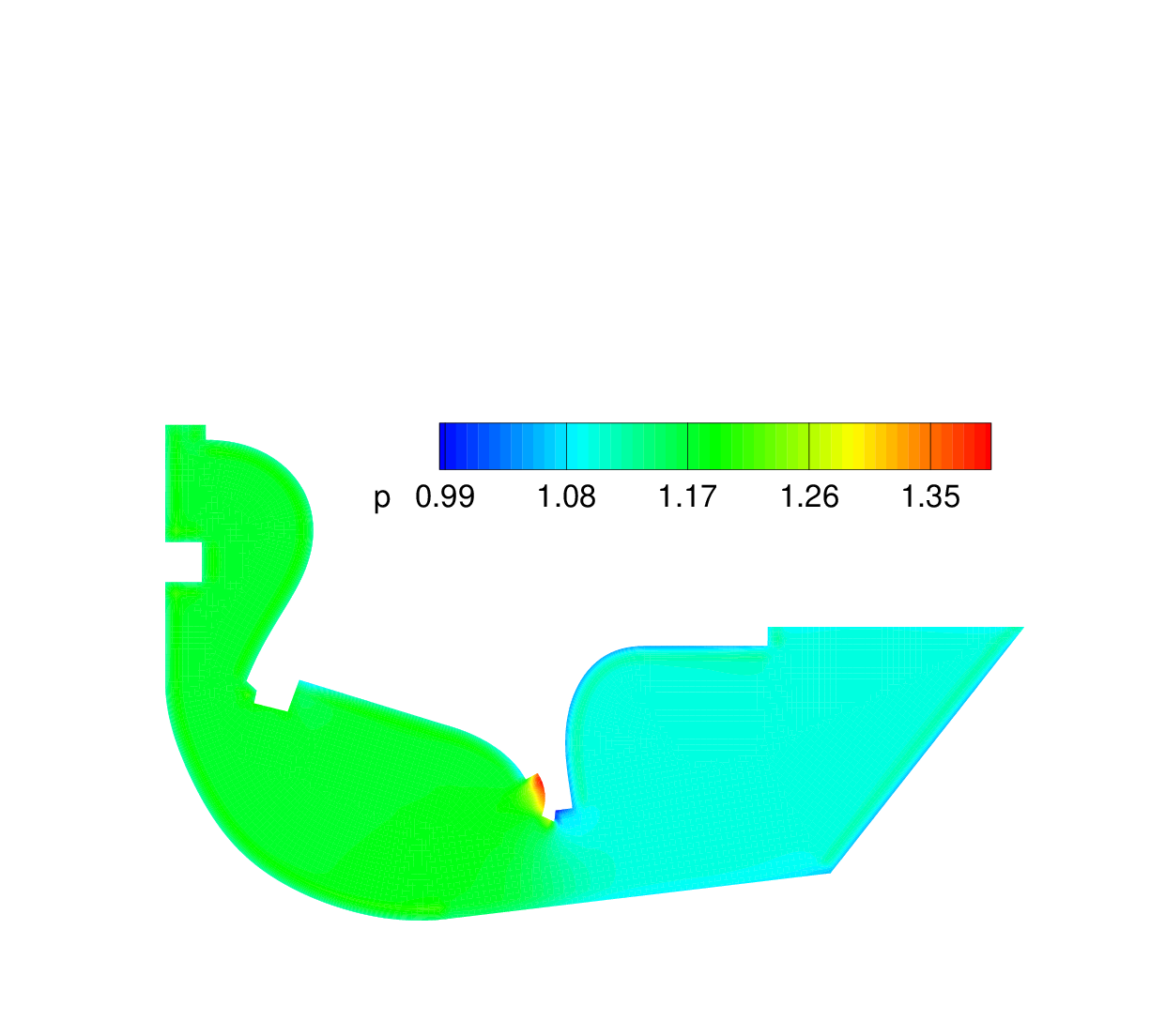}}
    {\includegraphics[trim={0 0 50 200},clip,width=0.32\linewidth]{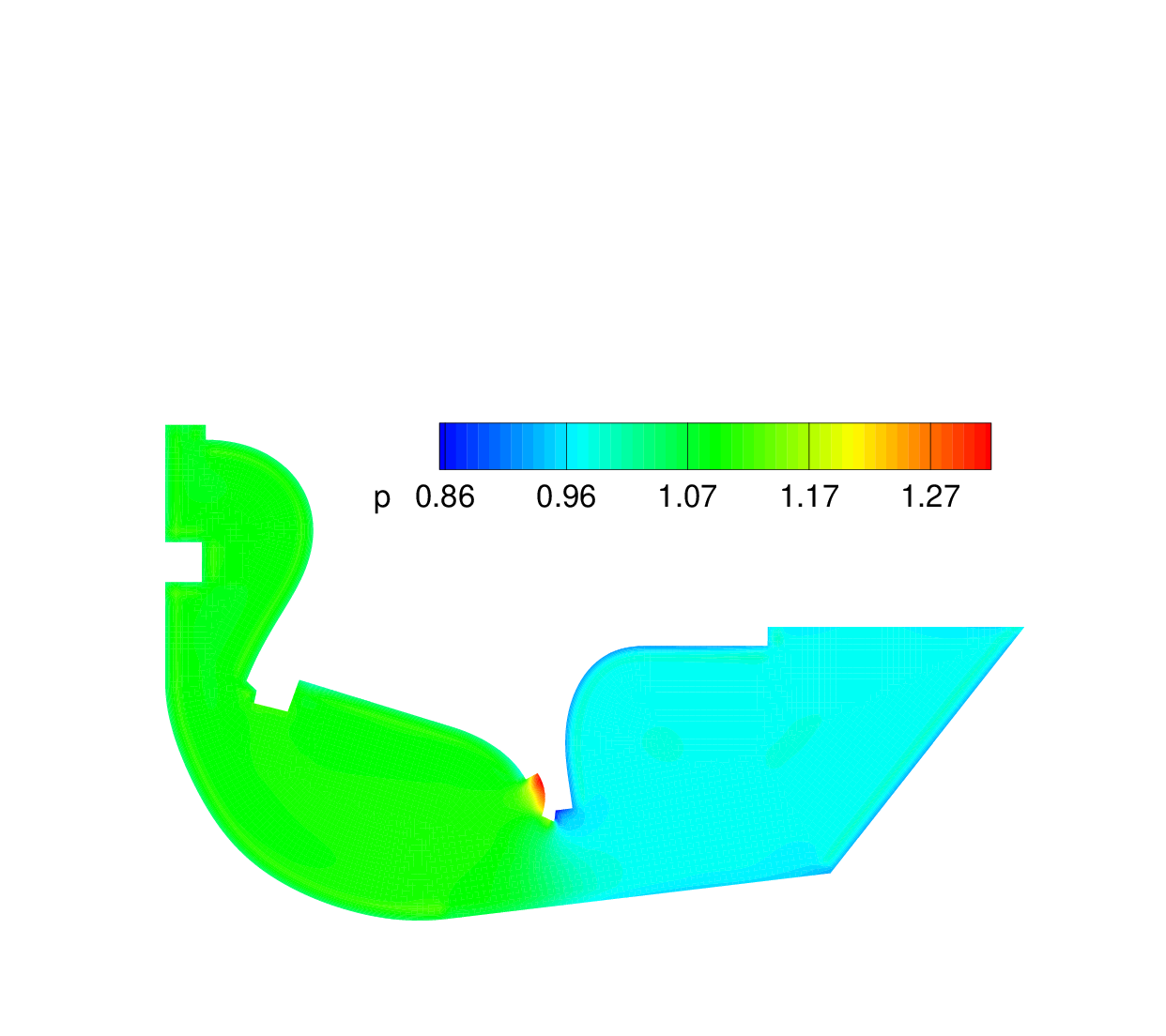}}
    {\includegraphics[trim={0 0 50 200},clip,width=0.32\linewidth]{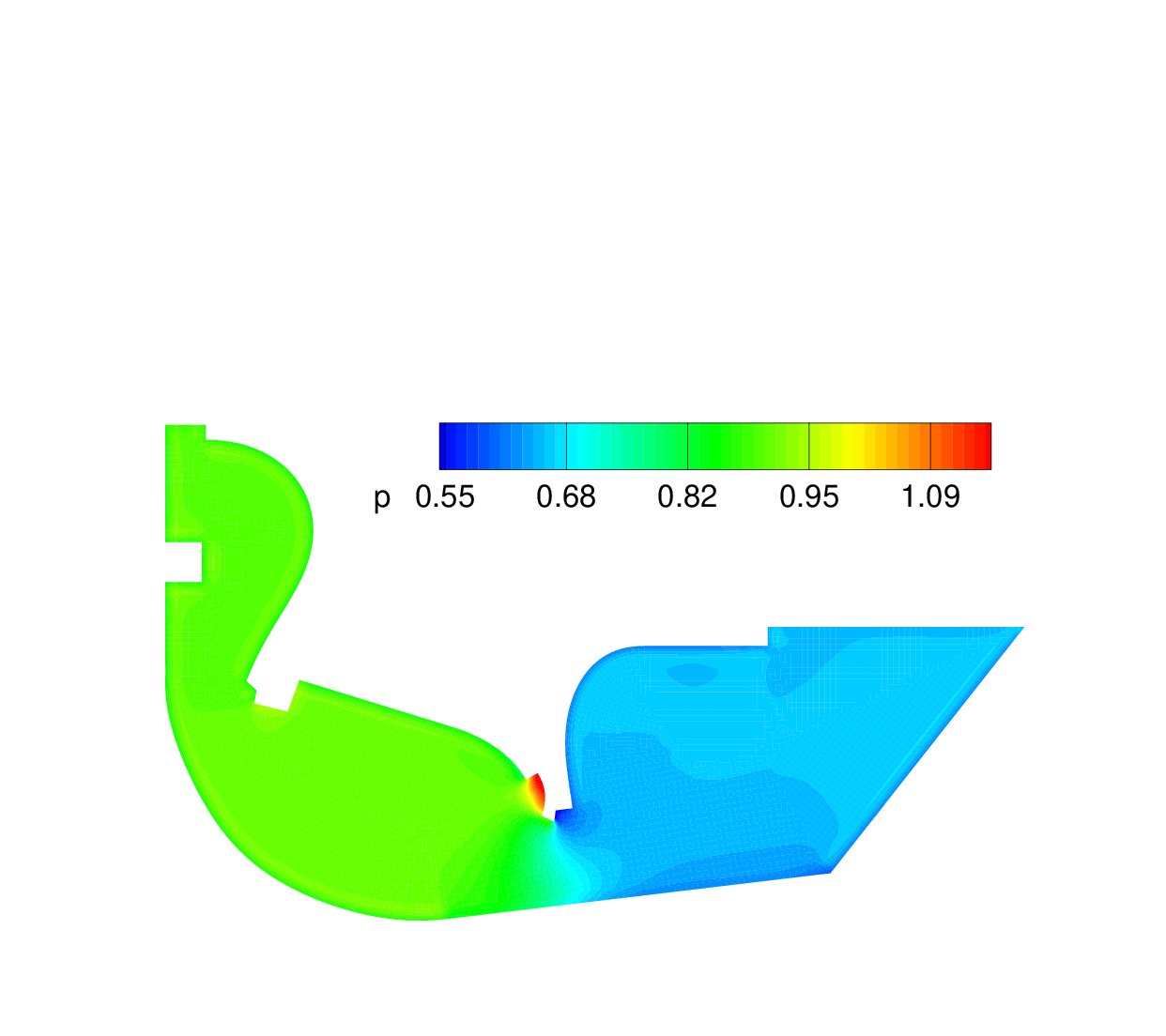}}\\
     {\includegraphics[trim={50 150 50 100},clip,width=0.32\linewidth]{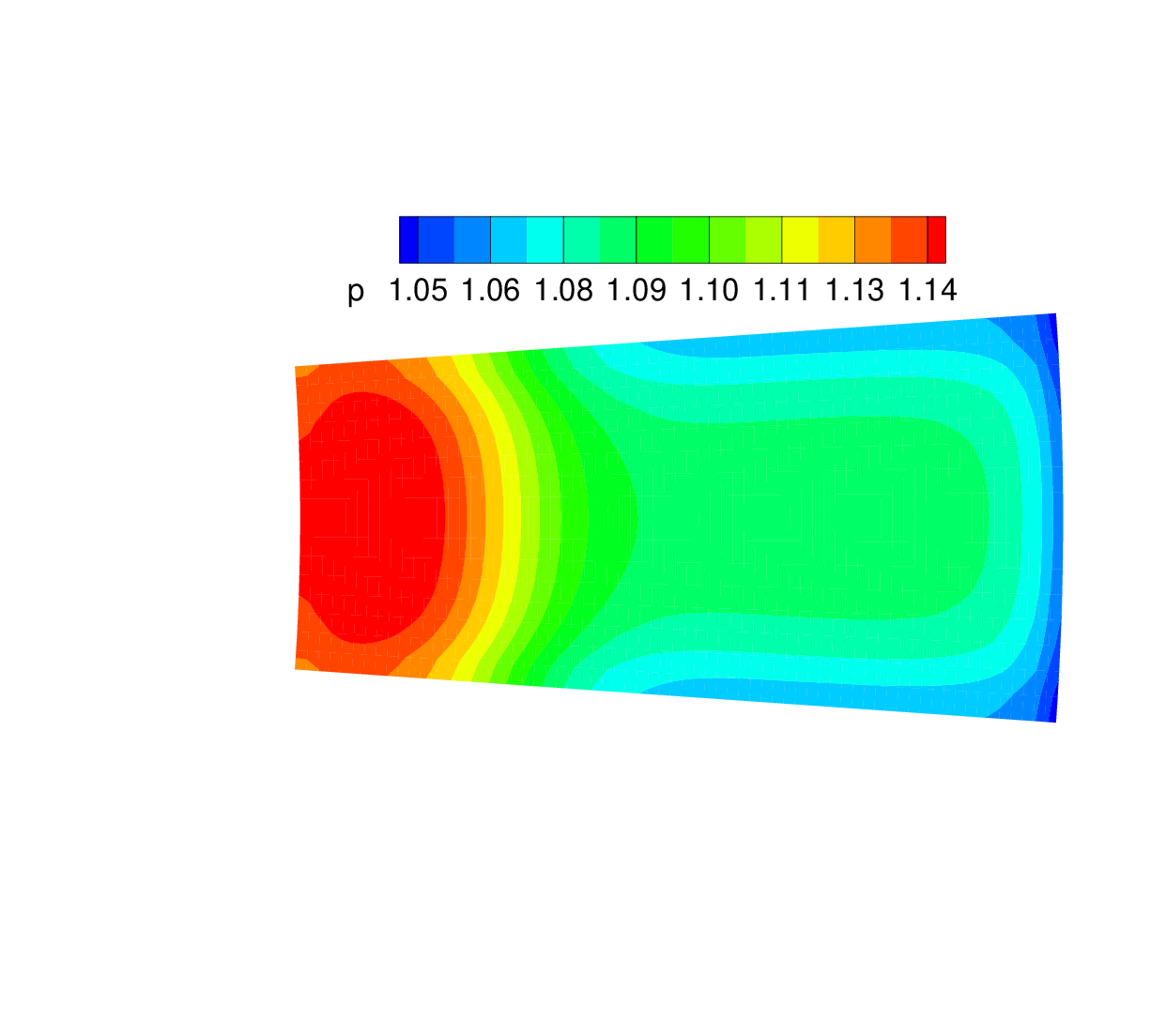}}
    {\includegraphics[trim={50 150 50 100},clip,width=0.32\linewidth]{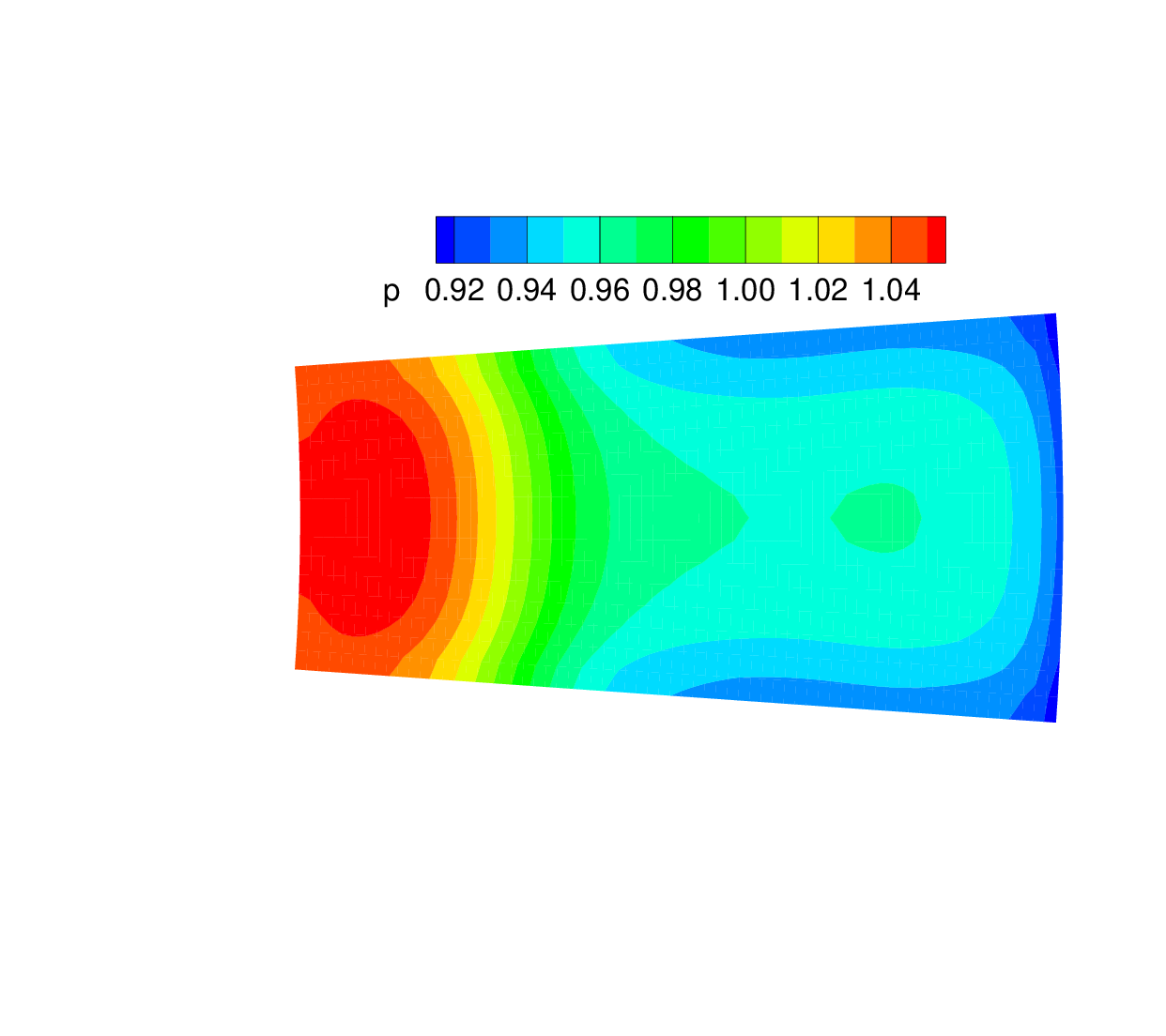}}
    {\includegraphics[trim={50 150 50 100},clip,width=0.32\linewidth]{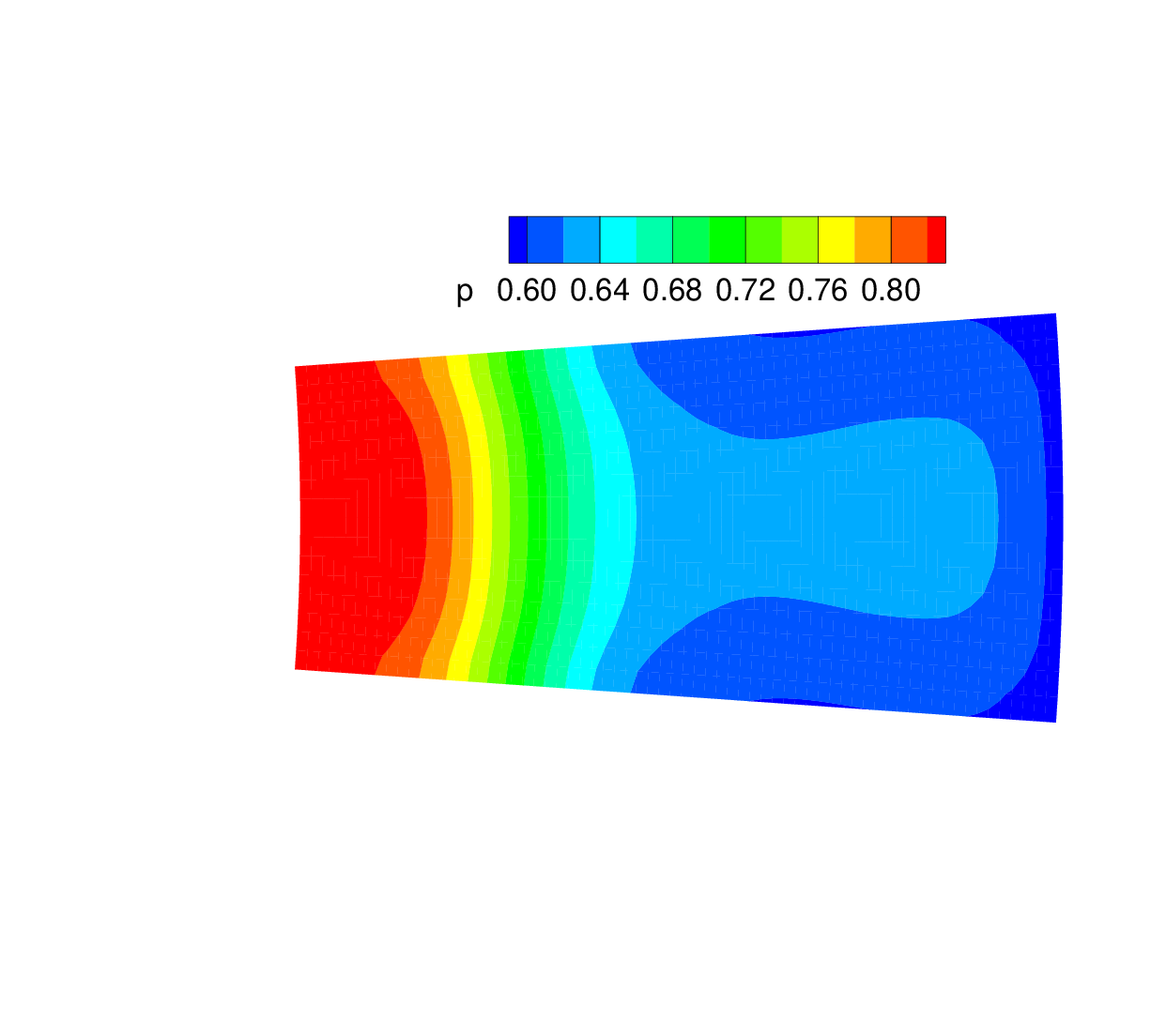}}
    \caption{Temperature and pressure  distributions at the midsection ($z=0$) and pump opening under different absorptivity when Kn = 0.054. The inlet gas temperature is 1000 K, while the wall temperature is 300~K.
    }
    \label{Kn0.054}
\end{figure}

\subsection{Role of inlet temperature}

We now raise the gas temperature to \leir{$T_0=1000$~K}, while keep the wall temperature at 300~K. 
Figure~\ref{Kn0.054} depicts the spatial distribution of temperature and pressure. Significant temperature differences are observed near both the inner entry gap and the outer entry gap, highlighting the complex thermodynamic phenomena within the system and the evolution of temperature gradients across different regions.
Figure~\ref{Kn0.054} also shows the pressure distribution at the outlet pump. The pressure profile manifests itself in a strip-like configuration that stretches from left to right, with a noticeable disparity observed between approximately one third of the area on the left side compared to the right side. Notably, a higher gas density emanates from the left side of the outlet pump, indicating a nonuniform distribution of pressure across the outlet. 

\begin{figure}[t]
    \centering
    \subfloat[$\zeta=0.05$]{\includegraphics[trim={0 0 0 50},clip,width=0.4\linewidth]{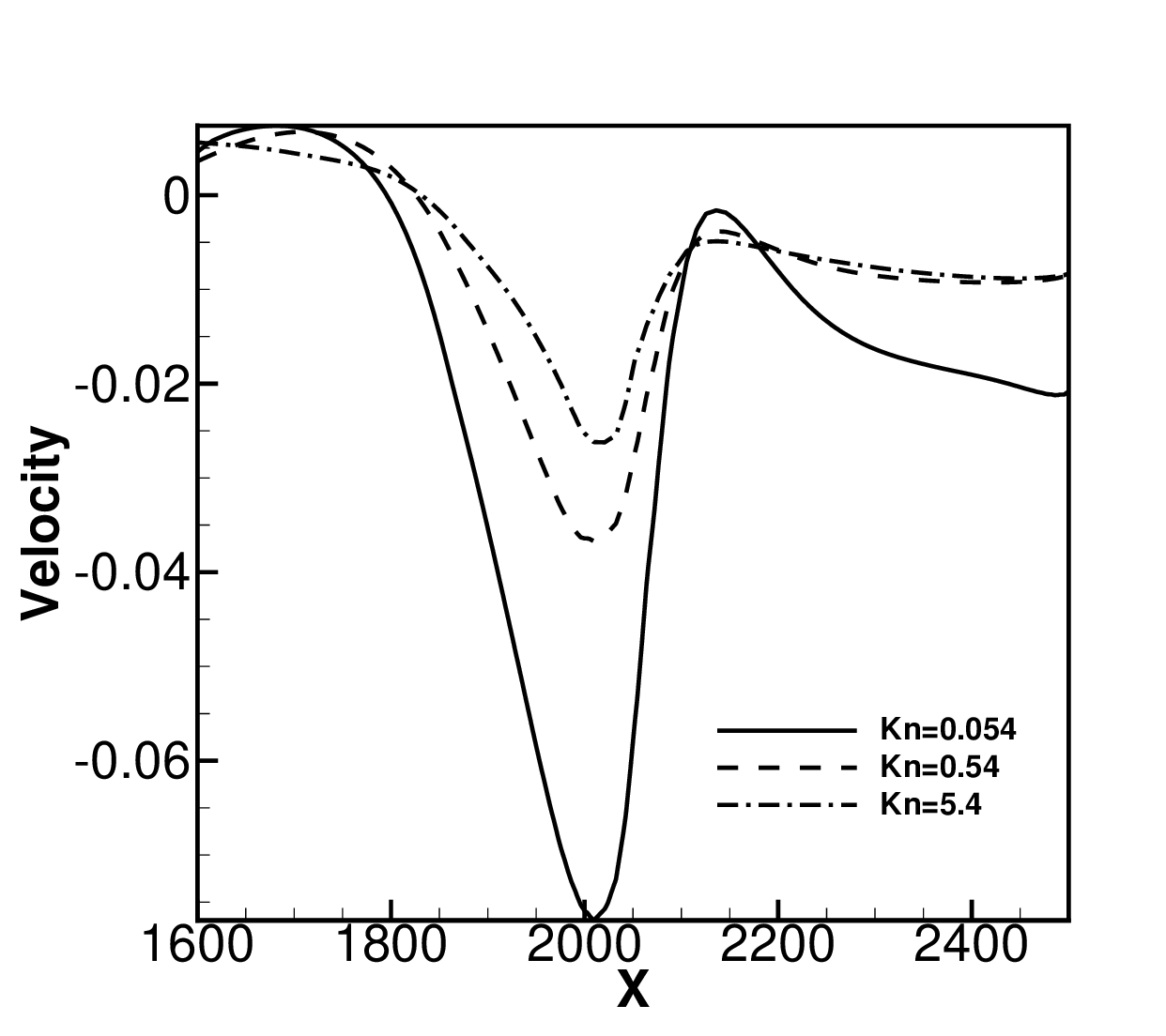}}
    \subfloat[$\zeta=0.1$]{\includegraphics[trim={0 0 0 50},clip,width=0.4\linewidth]{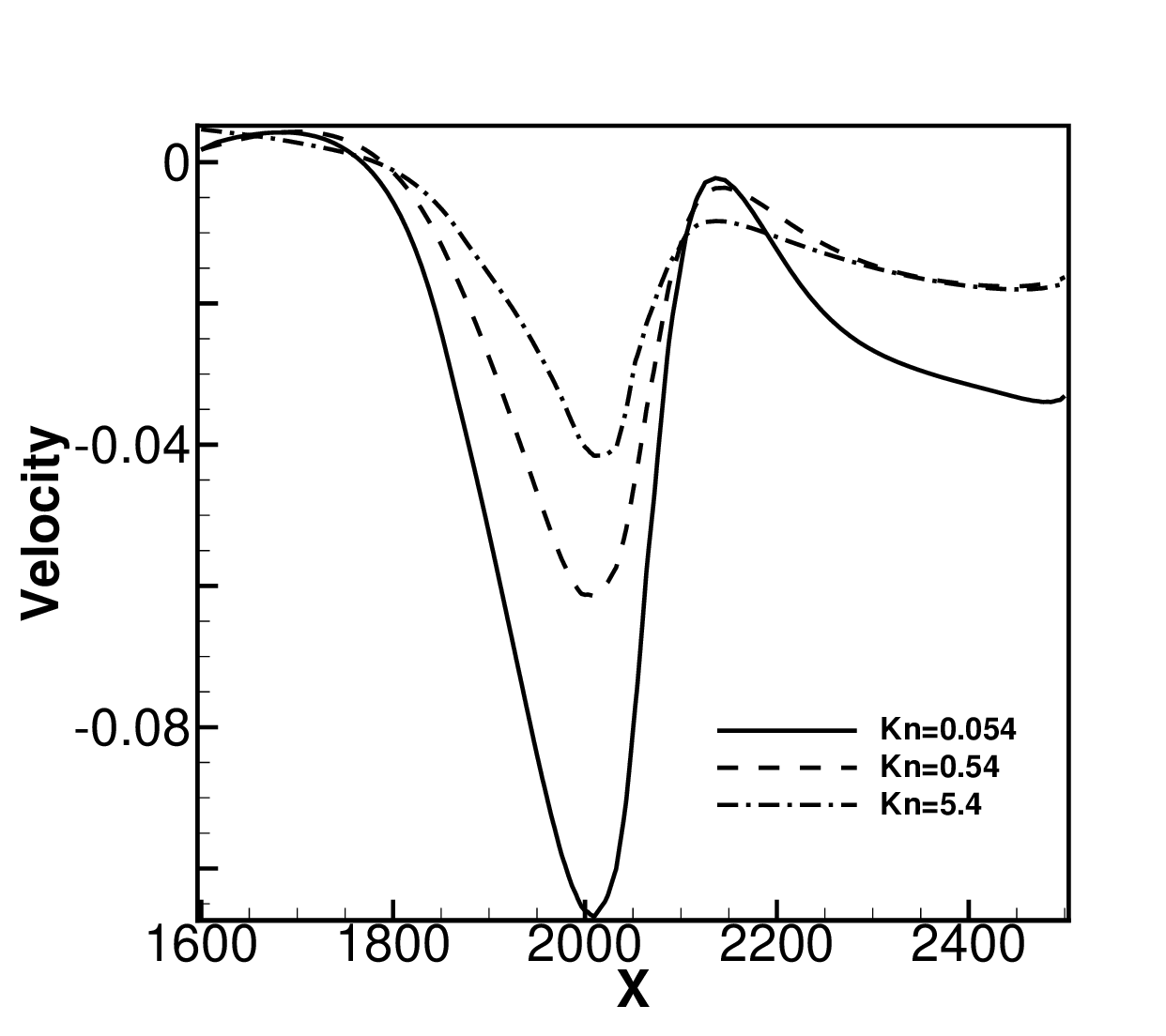}}\\
    \subfloat[$\zeta=0.3$]{\includegraphics[trim={0 0 0 50},clip,width=0.4\linewidth]{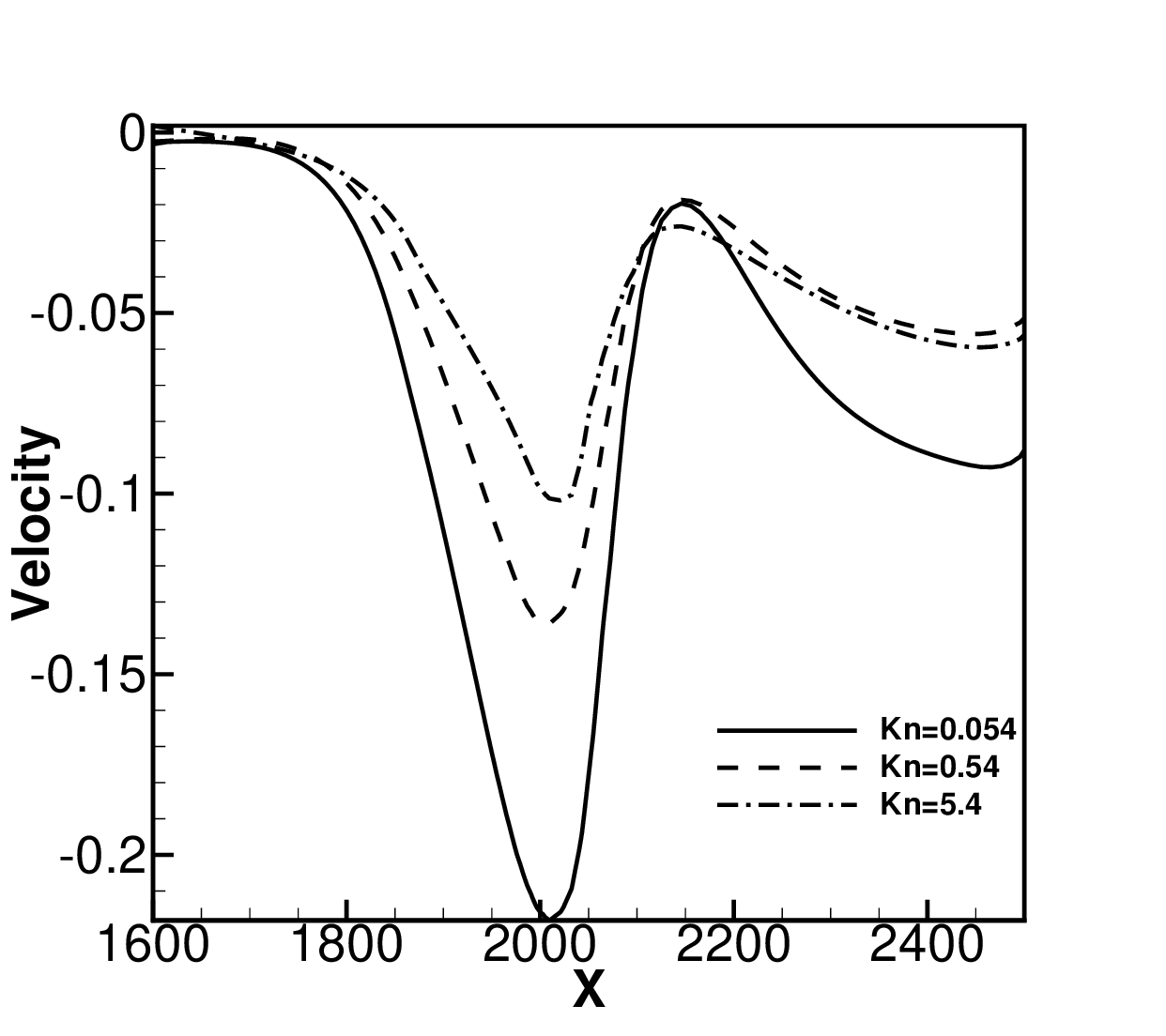}}
     \subfloat[Mass flow rate]{\includegraphics[trim={0 0 0 50},clip,width=0.4\linewidth]{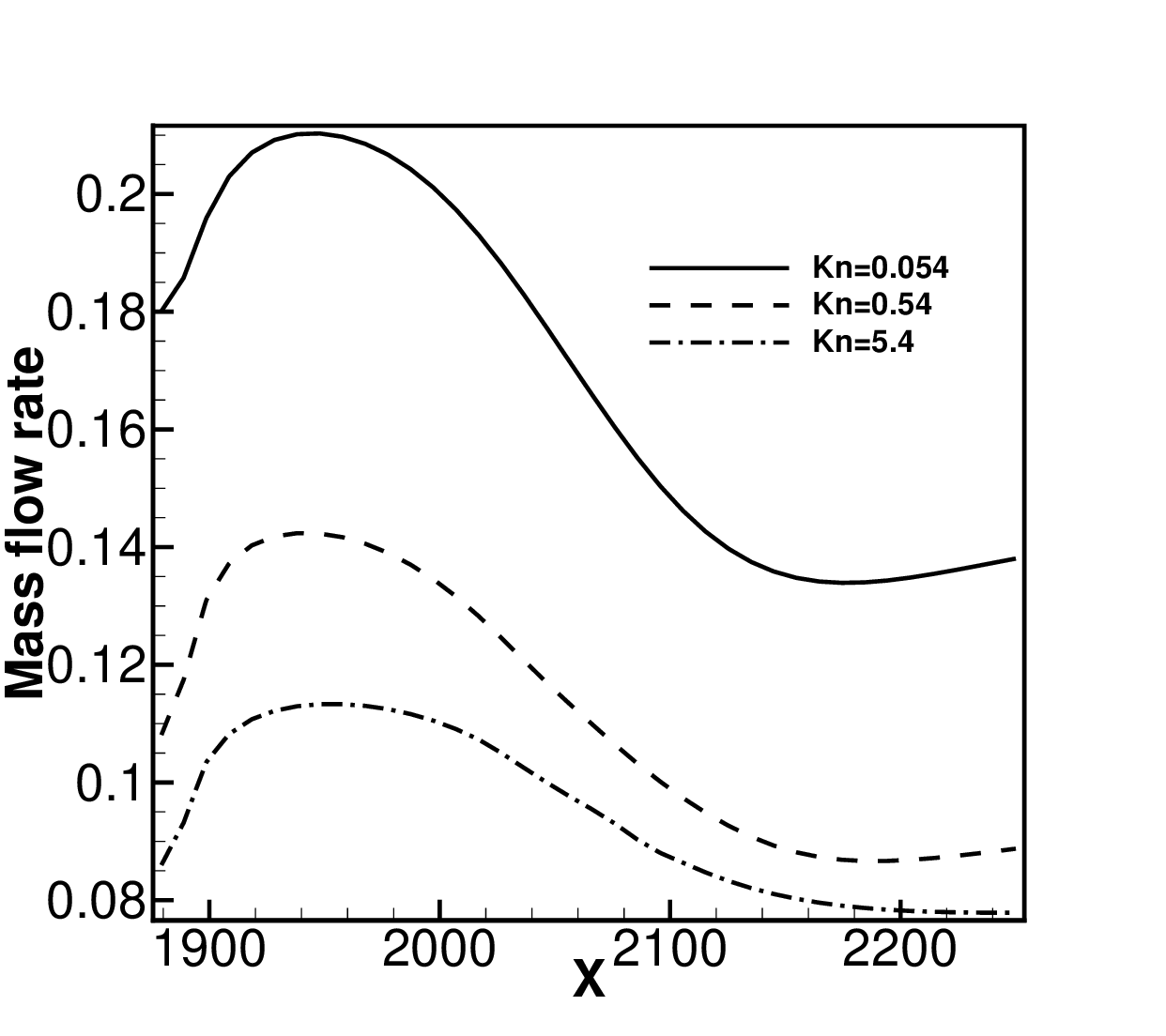}}
    \caption{
    (a-c) The velocity in the y-direction of axial section. (d) Mass flow distribution along the central axis ($z = 0$) of the absorption pump with absorptivity = 0.3.
    The inlet gas temperature is 1000~K, while the wall temperature is 300~K. 
    }
    \label{054_velocity}
\end{figure}

Figure~\ref{054_velocity}(a-c) shows the velocity profile $u_y$ in the midsection ($z=0$). Compare to Fig.~\ref{zhong_velocity},  the speed also reaches its maximum velocity around $x=2000$, and the velocity increases with the increase of gas temperature. Also, the velocity becomes much more sensitive to the Knudsen number.
Figure~\ref{054_velocity}(d) shows the mass flow rate along the central axis ($z=0$) of the absorption pump for different Knudsen numbers. The peak mass flow rate appears around $x=1940$, rather than at $x=2000$ where the gas speed is maximum. This is because the gas density at the pumping opening drops significantly as $x$ increases, see the third row in Fig.~\ref{Kn0.054}. Compared to Fig.~\ref{zhong_velocity}(d), the mass flow rate varies significantly with the Knudsen number, indicating the significant impact of gas temperature on mass flow rate. 



\begin{figure}[t!]
    \centering
    \subfloat[]{\includegraphics[trim={10 0 30 60},clip,width=0.45\linewidth]{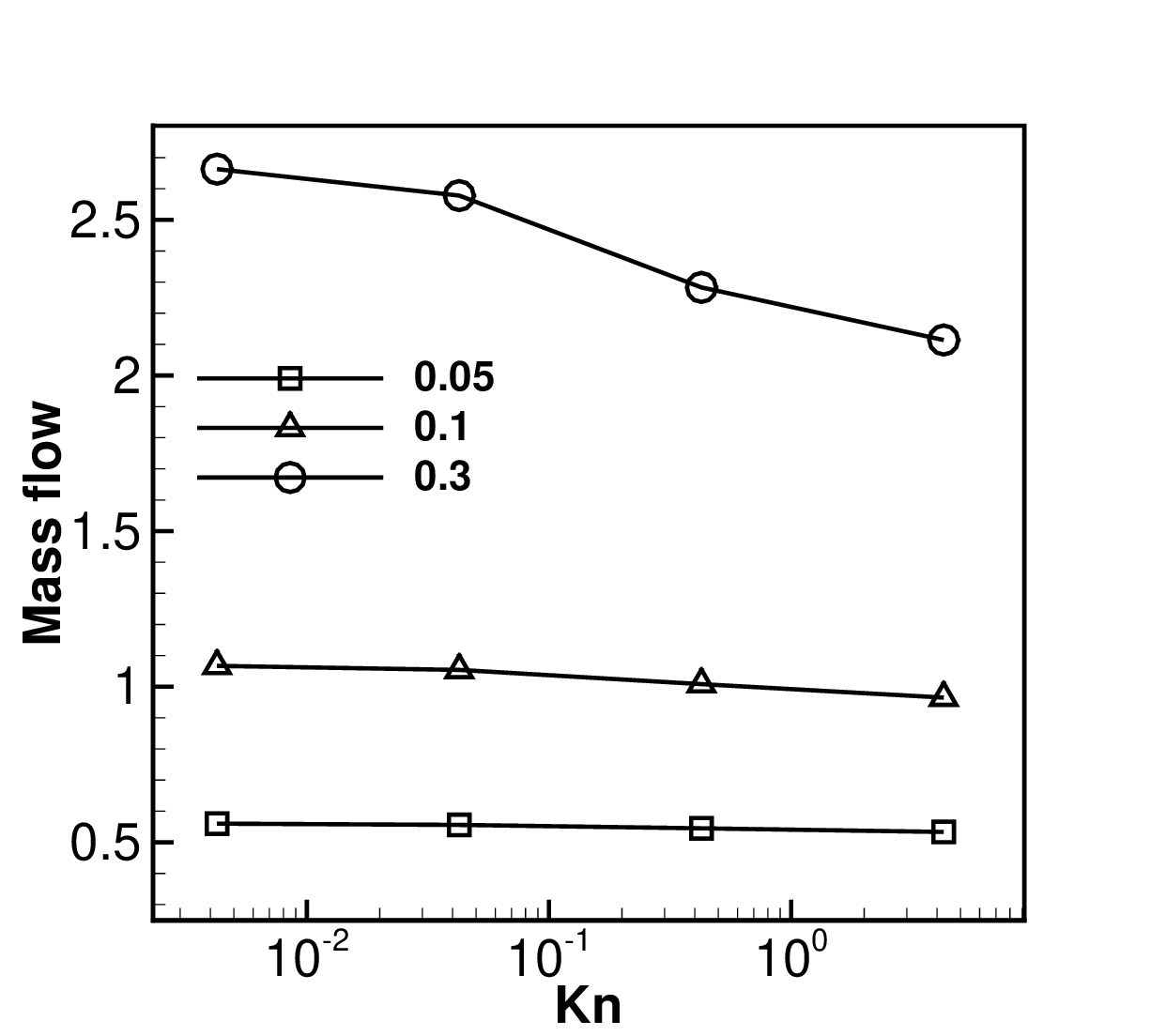}}
    \subfloat[]{\includegraphics[trim={10 0 30 60},clip,width=0.45\linewidth]{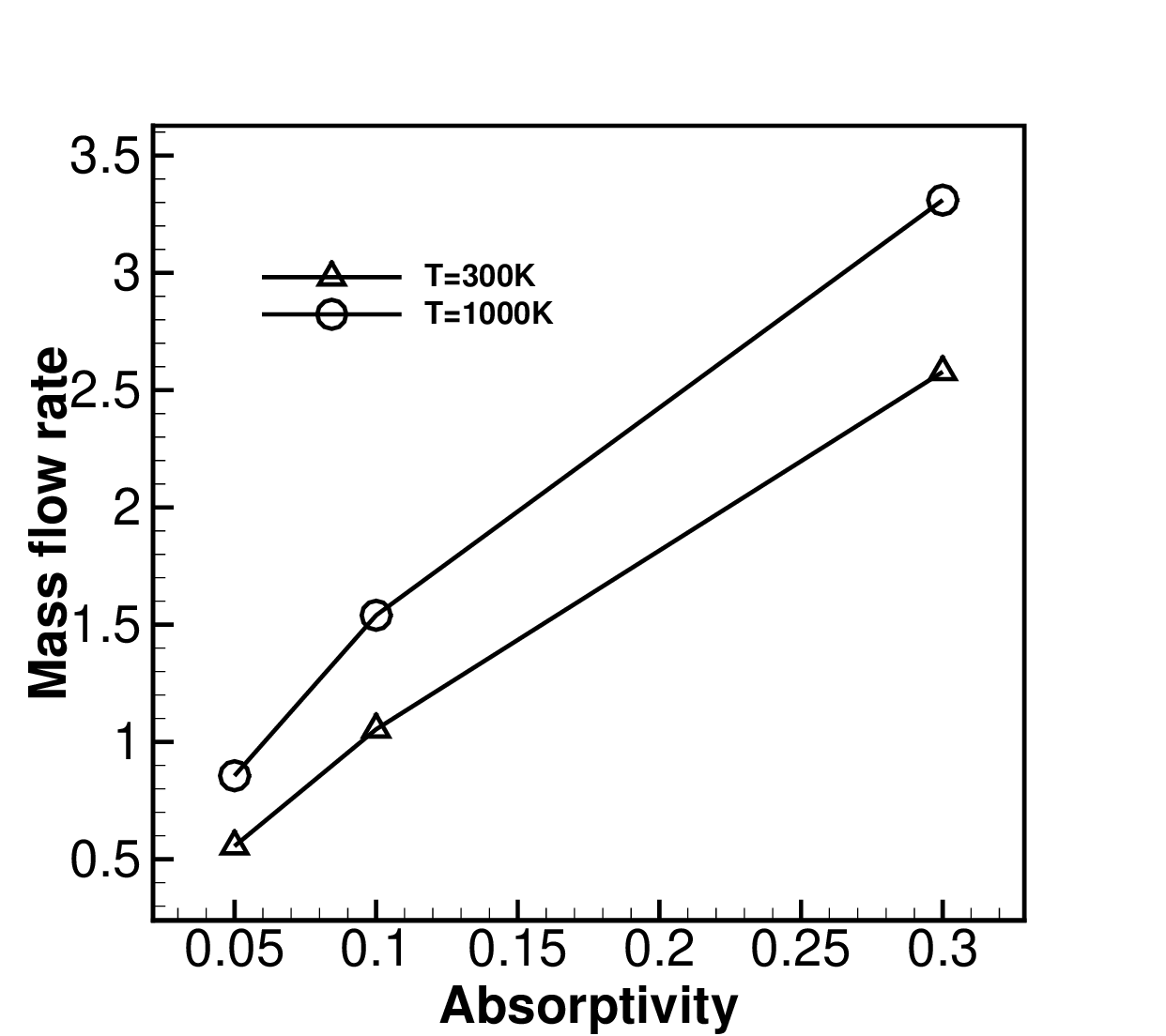}}\\
    \subfloat[]{\includegraphics[trim={10 0 30 60},clip,width=0.45\linewidth]{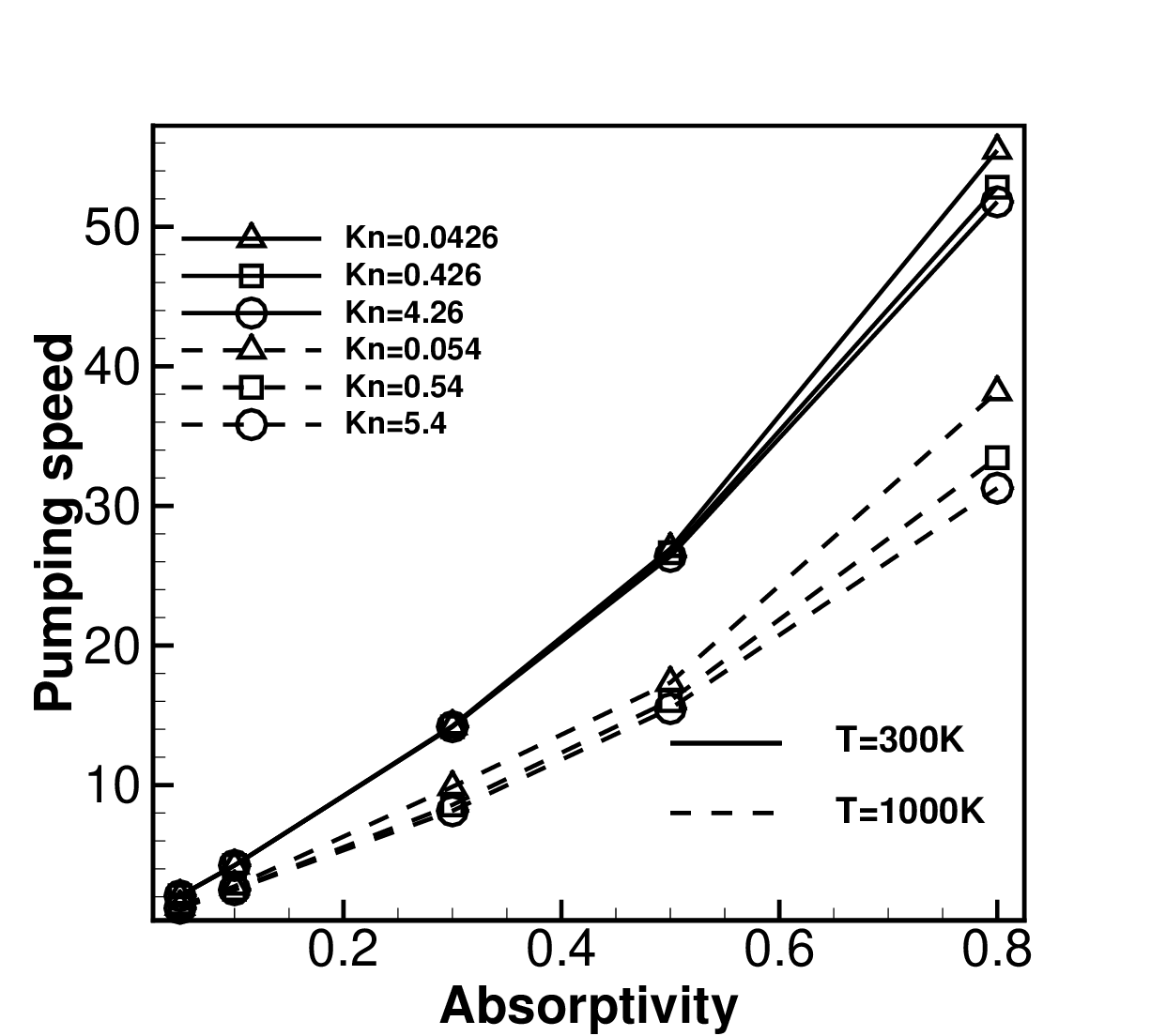}}
    \caption{ 
    (a) The mass flow under different absorptivity and Knudsen number when gas temperature is 300~K. (b) The mass flow under different absorptivity and temperature when Kn = 0.0426. (c) Pumping speed under different absorptivity and Knudsen number.
    }
    \label{mass_flow}
\end{figure}

\subsection{The pumping speed}

The pumping speed is a important parameter in the design of divertor, which is defined as the mass flow rate over the average gas density:
\begin{equation}
    S_p( \zeta ) =\frac{\text{mass flow rate}}{\text{average density}}= \frac{\int_S \rho \bm{u}\cdot\bm{n} dS }{(\int_S\rho dS)/S},
\end{equation}
where $\bm{v}$ is the velocity of the gas, $S$ is the area of the pump, and $\bm{n}$ is the outward normal vector of the pump surface. 

In Fig.~\ref{mass_flow}(a), a direct correlation is observed between the mass flow rate through the absorption pump and the Knudsen number at various absorptivity. The results indicate that as the Knudsen number increases, the gas within the deflector tends to become rarefied, leading to a gradual decrease in the mass flow rate through the absorption pump. This trend is further accentuated with higher absorptivity. However, the pumping rate shows minimal variation with changes in the Knudsen number when absorptivity is small. This suggests that the efficiency of the absorption pump is significantly influenced by the rarefaction of the gas within the deflector, particularly under the conditions of increased Knudsen numbers and absorptivity. The findings underscore the importance of considering the Knudsen number and absorptivity in optimizing the performance of absorption pumps in rarefied gas flow environments.

Figure~\ref{mass_flow}(b) shows the mass flow through the absorption pump under different absorptivity and temperature. With an increase in absorptivity, the mass flow rate showed a trend of logarithmic increase, and the increase of the temperature significantly increased the mass flow rate through the absorption pump.

Figure~\ref{mass_flow}(c) shows the correlation between velocity and absorptivity at various Knudsen numbers. It is seen that for higher absorptivity, there is a slight reduction in velocity with an increase in the Knudsen number. At elevated temperatures, the mass flow rate of the outlet pump increases, while the pumping speed decreases. This indicates that the impact of temperature elevation on density distribution is more pronounced.

\section{Conclusions} \label{section_summary}

We have simulated a simplified version of the latest three-dimensional divertor using a general synthetic iterative scheme, with the velocity space discretized in cylindrical coordinates. Due to its asymptotic-preserving nature and rapid convergence, and given that the gas flow inside the divertor is much lower than the sound speed, the simulation efficiency of GSIS is significantly higher than that of the traditional DSMC method.
Using this efficient numerical method, we have analyzed the flow fields inside the divertor across a wide range of Knudsen numbers, absorptivity levels, and gas temperatures. The simulation results indicate that higher absorption rates lead to a substantial increase in mass flow rate and pumping speed through the absorption pump. Additionally, increasing the gas temperature also significantly enhances the mass flow rate.
The numerical method and findings provide valuable tool and insights into the performance of the new Tokamak model under various conditions, supporting the ongoing development of nuclear fusion technology.

\section*{Acknowledgments}

This work is supported by the "Climbing Program" for scientific and Technological Innovation in Guangdong (pdjh2024c10701).

\section*{Declaration of interests} 
The authors report no conflict of interest.


\bibliography{thesisBib}

\end{document}